\documentstyle[aaspp4,epsf,graphicx]{article}

\begin{document}

\title{Multiwavelength observations of Mkn 501 during the 1997 high state}

\author{ D. Petry\altaffilmark{1}, M. B\"ottcher\altaffilmark{2,3},
V. Connaughton\altaffilmark{4}, A. Lahteenmaki\altaffilmark{5},
T. Pursimo\altaffilmark{5}, C.M. Raiteri \altaffilmark{6},
F. Schr\"oder\altaffilmark{7}, A. Sillanp\"a\"a\altaffilmark{5},
G. Sobrito\altaffilmark{6}, L. Takalo\altaffilmark{5},
H. Ter\"asranta\altaffilmark{8}, G. Tosti\altaffilmark{9},
and M. Villata\altaffilmark{6} 
}
\altaffiltext{1}{Institut de F\'{\i}sica d'Altes Energies, 
Universitat Autonoma de Barcelona, 08193 Bellaterra, Spain}
\altaffiltext{2}{Space Physics and Astronomy Department, 
Rice University, MS 108, 6100 S. Main Street, Houston, TX 77005 - 1892, USA}
\altaffiltext{3}{Chandra Fellow}
\altaffiltext{4}{Marshall Space Flight Center, Alabama, USA (National Research Council Fellow)}
\altaffiltext{5}{Tuorla Observatory, 21500 Piikki\"o, Finland}
\altaffiltext{6}{Osservatorio Astronomico di Torino, 
Strada Osservatorio 20, 10025 Pino Torinese, Italy}
\altaffiltext{7}{Universit\"at Wuppertal, Fachbereich Physik, 
Gau\ss{}-Str.20, 42119 Wuppertal, Germany}
\altaffiltext{8}{Mets\"ahovi Radio Observatory, Mets\"ahovintie 114, 
    02540 Kylm\"al\"a, Finland}
\altaffiltext{9}{Osservatorio Astronomico di Perugia, Via Bonfigli, 06123 Perugia, Italy}

\bigskip
\centerline{\it Submitted to The Astrophysical Journal}
\bigskip

\begin{abstract}

During the observation period 1997, the nearby
Blazar Mkn~501 showed extremely strong emission
and high variability. We examine multiwavelength
aspects of this event using radio, optical, soft 
and hard X-ray and TeV data. We concentrate on the 
medium-timescale variability of the broadband 
spectra, averaged over weekly intervals. 

We confirm the previously found correlation between
soft and hard X-ray emission and the emission at
TeV energies, while the source shows only minor
variability at radio and optical wavelengths. The
non-linear correlation between hard X-ray and TeV 
fluxes is consistent with a simple analytic estimate 
based on an SSC model in which Klein-Nishina effects 
are important for the highest-energy electrons in the
jet, and flux variations are caused by variations of
the electron density and/or the spectral index of the
electron injection spectrum.

The time-averaged spectra are fitted with a 
Synchrotron Self-Compton (SSC) dominated
leptonic jet model, using the full Klein-Nishina cross
section and following the self-consistent evolution of
relativistic particles along the jet, accounting for 
$\gamma\gamma$ absorption and pair production within 
the source as well as due to the intergalactic infrared 
background radiation. The contribution from external
inverse-Compton scattering is tightly constrained by
the low maximum EGRET flux and found to be negligible
at TeV energies. 
We find that high levels of the X-ray and TeV fluxes 
can be explained by a hardening of the energy spectra 
of electrons injected at the base of the jet, in 
remarkable contrast to the trend found for 
$\gamma$-ray flares of the flat-spectrum radio 
quasar PKS~0528+134. 

\keywords{BL Lacertae objects: individual: Mkn 501 ---
radiation mechanisms: non-thermal}

\end{abstract}

\section{Introduction}

The BL Lac object Mkn 501 is very close ($z = 0.0337$, Ulrich 
et al. \cite{ulrich}) and has  been studied extensively 
at all wavelengths. Together with its sister object Mkn 421, 
it was among the two first BL~Lac objects with known radio 
(Colla et al. \cite{colla}), X-ray (Schwartz et al. 
\cite{schwartz}) and TeV gamma-ray (Quinn et al. \cite{quinn}, 
Bradbury et al. \cite{bradbury}) counterparts. Recently, it was
also marginally detected at photon energies $> 100$~MeV by the 
EGRET instrument on board CGRO (Kataoka et al. \cite{kataoka}).

During 1997,  the object was found to be in an
extreme high state with a TeV flux on average 20 times higher than
in 1996 (Breslin et al. \cite{breslin}). The source  exhibited strong 
variability on timescales of days with a possible quasi-periodically 
varying component with a timescale of about 25 days (Kranich et al. \cite{kranich}). 
To complete the list of reasons for excitement for the observers, 
Mkn~501 reached in some of its flares fluxes of more than 
10$^{-10}$~cm$^{-2}$s$^{-1}$ (above 1.5~TeV) - the most intense 
TeV emission ever measured so far from any astronomical object. 
However, the shortest observed variability timescale (5 hours, 
Aharonian et al. \cite{hegraparti}) was significantly longer 
than that observed for Mkn~421 (Gaidos et al. \cite{gaidos}).

All available TeV observatories monitored the event for several 
months (Samuelson et al. \cite{whipple}, Aharonian et al. 
\cite{hegraparti},\cite{hegrapartii}, Hayashida 
et al. \cite{telarray}, Djannati-Ata\"{\i} et al. \cite{cat}). The most complete dataset 
was produced by the HEGRA Cherenkov Telescope ``1'' (CT1) 
(Aharonian et al., \cite{hegrapartii}) which was even able to 
observe Mkn~501 under the presence of moonlight, however with 
reduced sensitivity, thereby filling many gaps in the lightcurve. 
This telescope also obtained the confirmatory observations in 1996.

Also from HEGRA come probably the most accurate spectral measurements
in the TeV regime. They were carried out by the HEGRA system of (at 
the time 4) Cherenkov telescopes (CTS) and are largely concurrent with
the CT1 measurements, however with less time coverage (Aharonian et al. 
\cite{hegraparti}).

The origin of the TeV $\gamma$-ray emission and the reasons for its 
variability are still essentially unknown. The most popular models 
explain the TeV emission as near-infrared to UV photons which have 
been upscattered via the inverse Compton effect by very high energy 
electrons which are known from radio, optical and X-ray observations 
to be present in the jets of BL Lac objects. Possible sources of the 
seed photons for Compton scattering are the synchrotron radiation produced 
within the jet by the same population of relativistic electrons (synchrotron 
self-Compton model, SSC; Marscher \& Gear \cite{marscher}; Maraschi, 
Celotti \& Ghisellini \cite{maraschi}; Bloom \& Marscher \cite{bloom}), 
or radiation from outside the jet (external inverse Compton model, EIC). 
This external radiation could be the quasi-thermal radiation field of 
an accretion disk surrounding a supermassive black hole which is 
generally believed to power the relativistic jets. The accretion 
disk radiation can enter the jet either directly (Dermer, Schlickeiser 
\& Mastichiadis \cite{dermer92}, Dermer \& Schlickeiser \cite{dermer93}) 
or after being rescattered by circumnuclear material (Sikora, Begelman 
\& Rees \cite{sikora}, Blandford \& Levinson \cite{blandford}, Dermer, 
Sturner \& Schlickeiser \cite{dermer97}). It is also possible that 
synchrotron radiation produced within the jet and reflected by 
circumnuclear debris is the dominant source of soft photons during 
flares (Ghisellini \& Madau \cite{ghisellini96}, Bednarek 
\cite{bednarek}), although it has been shown that this process 
is unlikely to be efficient in the case of BL~Lac objects 
(B\"ottcher \& Dermer \cite{boettcher98}).

As an alternative, Mannheim (\cite{mannheim93}) has 
suggested a hadronic model in which protons are the 
primarily accelerated particles in the jet and the 
$\gamma$-ray emission is produced by secondary pions 
and electron-positron pairs produced in photopion and 
photopair production interactions of the ultrarelativistic 
protons in the jet with external radiation. The time-averaged
broadband spectrum of Mkn~501 has been fitted using this model
by Mannheim et al. (\cite{mannheim96}, \cite{mannheim98}). 
However, the attempts to explain the short and intermediate-term 
variability of blazars with hadronic jet models have only just started
(Rachen \& Mannheim \cite{rachen99}). 
For this reason, we concentrate on leptonic jet models in this paper 
since the variability time scales predicted by these models 
are in good agreement with the observed intraday variability 
of blazars.

Recently, Fossati et al. (\cite{fossati}) have compared the 
broadband spectra of different types of $\gamma$-ray emitting 
AGN and suggested a continuous sequence FSRQs (flat spectrum 
radio quasars) $\to$ LBLs (low-frequency peaked BL~Lac objects) 
$\to$ HBLs (high-frequency peaked BL~Lac objects), characterized 
by decreasing bolometric luminosity, decreasing dominance of the 
total energy output in $\gamma$-rays compared to the emission at 
lower frequencies and a shift of the peak frequencies of the 
synchrotron and the $\gamma$-ray component towards higher 
frequencies. Recent modelling efforts of various blazar-type 
AGN have revealed that this sequence is consistent with a 
decreasing importance of external radiation as a source of 
soft photons for Compton scattering in the jet (Ghisellini 
et al. \cite{ghisellini98}). This suggests that the extreme 
HBLs like Mkn~501 or Mkn~421 can be well fitted with strongly 
SSC-dominated jet models, as was shown, e. g., by Mastichiadis 
\& Kirk (\cite{mastichiadis97}) and Pian et al. (\cite{pian}).

\begin{figure}
\epsfysize=21cm
\epsffile{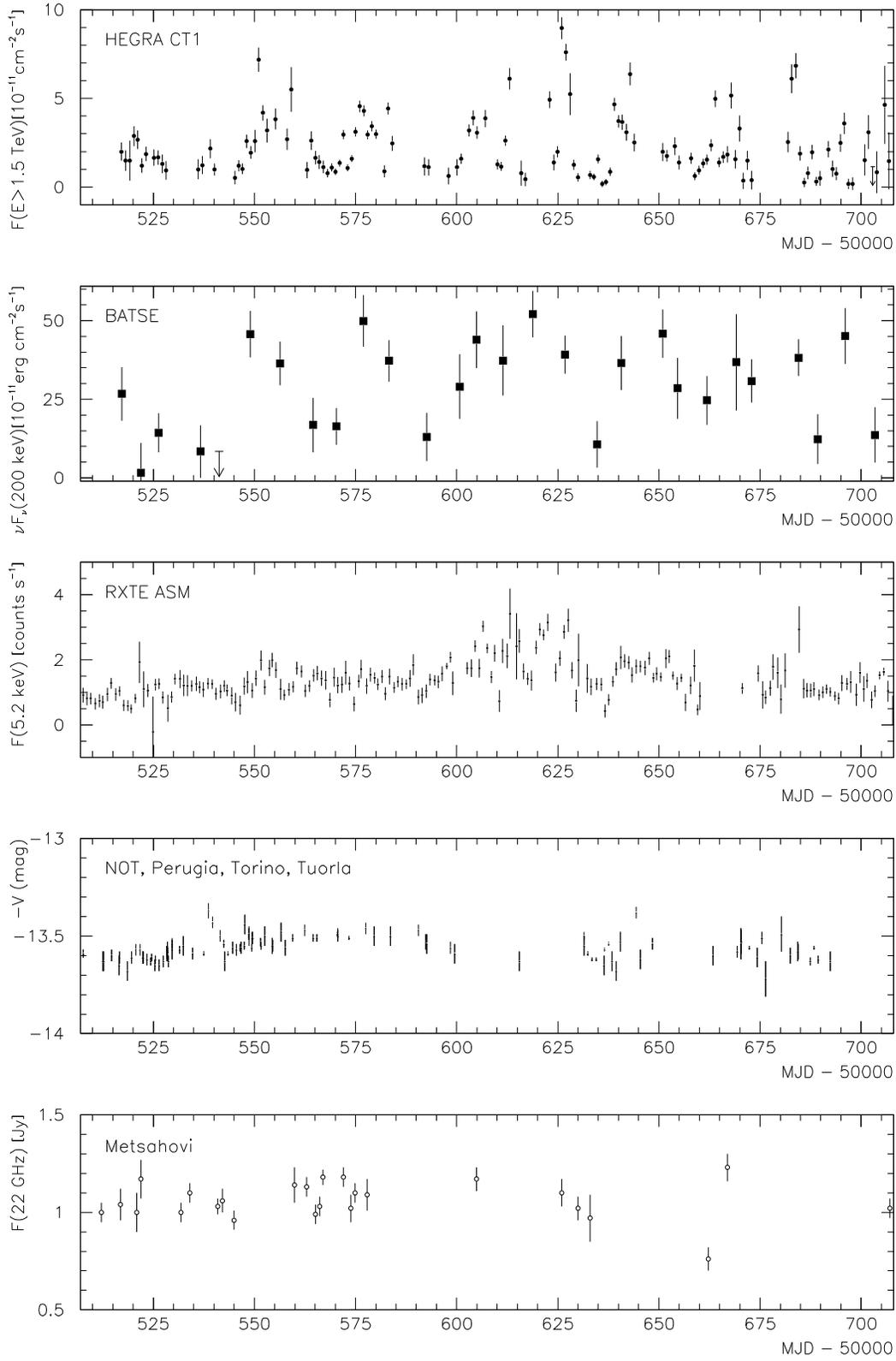}
\figcaption{\label{fig-lightcurve} 
The lightcurve data which were used in this paper to construct
weekly spectral energy distributions. 
See section \protect\ref{sec-observations} for references.}
\end{figure}

From the SSC model one expects a strong correlation between 
the X-ray and the TeV emission. And indeed, by comparing the 
daily TeV measurements with daily averages from the All Sky 
Monitor (ASM) on board the Rossi X-Ray Timing Explorer (RXTE), 
a significant correlation with a most probable time-lag of 0 
days (no time-lag) was found (Aharonian et al. 
\cite{hegraparti,hegrapartii}). Furthermore, a clear overall 
X-ray high state was visible in the 1997 ASM measurements which 
coincided with the TeV high state (see also figure \ref{fig-lightcurve}). 
And, as discovered by BeppoSAX, 
the X-ray peak of the spectral energy distibution (SED) had shifted 
from 10-20 keV in 1996 to 100-200 keV in 1997 (Pian et al. \cite{pian}), 
consistent with the assumption that the X-ray and TeV-$\gamma$-ray 
flares are produced by a more powerful electron acceleration, shifting 
the high-energy cutoff of the electron distribution to higher energies and
hardening the X-ray spectrum.
This  observation was, however, only made during one flare in 1997.
In this paper we give evidence that the hardening took place during
all flares in 1997.
An alternative explanation for the synchrotron spectral changes, in
terms of a steadily-emitting helical-jet model, has been presented by Villata
\& Raiteri (\cite{villata99}). 

Apart from the short-term variability on timescales of hours to days, 
there is also a longer-term variability in Blazars which has so far
been investigated mainly in the optical regime (see e.g. Katajainen et al. (\cite{katajainen}) 
and references therein) and recently also in X-rays (e.g. Mc Hardy \cite{mchardy}).
This variability shows remarkable amplitudes (e.g. 4.7 mag for Mkn 421
in the optical) and in the case of OJ 287 there is even evidence for
periodicity. 

In this respect Mkn 501 is not yet very well explored. 
Mkn 501 has been observed in the TeV regime since its discovery
in 1995. The source showed low emission close to the sensitivity
limits until the onset of the 1997 high state.
Since the duty cycle of the TeV observatories is only about 10 \%,
this does not prove the absence of strong short flares prior to this 
high state, but the
increased average intensity of the source can be described as
an increase in flaring probability from 1996 to 1997
by at least an order of magnitude. This description is especially
appropriate since
even during the high state, the source returns to quiescent
(comparable with 1996) levels of emission for periods of up to a few
days.  These transitions are seen irrespective of the presence of short,
strong flares on time-scales of a day.

The last observations in 1996 were made in August and found Mkn 501 still
quiescent while the first observations in 1997 were made in February 
and found the source already flaring. 
The transition from low to high flaring probability obviously
took place on timescales shorter than half a year. 
The correlated X-ray data from the RXTE ASM 
confirm this and show in addition that the change
has been a smooth process over several months (see e.g. figure 8 in
Aharonian et al. \cite{hegrapartii}).

In this article we explore the multi-wavelength variability 
of Mkn 501 over medium timescales, and try to relate its 
behaviour to the physical parameters of a leptonic jet model.  
For this purpose we construct weekly SEDs using the HEGRA CT1 
flux data and HEGRA CT System spectral data together with data 
from longer wavelengths, namely radio, optical, soft X-ray and 
hard X-rays. See table \ref{tab-instruments} for an enumeration 
of the instruments and energy ranges.

This paper may represent an important step toward understanding the 
origin of strong changes in the flaring probability of Mkn 501 
and Blazars in general, but our data is clearly not sufficient 
to give a final answer to this question.

All data except that from  HEGRA and the RXTE ASM are published 
here for the first time. The BATSE data are especially valuable 
since they confine the intensity at the X-ray peak of the SED.

\section{Observations and Data Analysis}
\label{sec-observations}

The observations used to fit the multi-wavelength spectrum cover
mostly the synchrotron part of the SED.  Only the TeV data explore what
is believed to be the Inverse Compton emission, although OSSE and EGRET
observations cover a few days in 1996 and 1997. As a guide line for our
fitting procedure we take into account the highest ever  observed EGRET 
flux from Mkn~501, $F(> 100 \, {\rm MeV}) = (32 \pm 13) \cdot 
10^{-8}$~photons~cm$^{-2}$~s$^{-1}$ with a photon spectral index
$\alpha = 1.6 \pm 0.5$, measured during 1996 Mar 25 \-- 28 (Kataoka
et al. \cite{kataoka}), as an upper limit.

In the definition of time-bins for the multi-wavelength dataset, 
we start by subdividing the HEGRA CT1 lightcurve and bin all other 
data accordingly. 

The data from HEGRA are binned in time from March through
October 1997  (MJD 50514 - MJD 50708) resulting in 28 equidistant time
bins.  This weekly temporal resolution is an order of magnitude larger
that the longest observed TeV intraday variability timescale of 15 h 
(Aharonian et al. \cite{hegraparti}) and is
believed appropriate, given the nature of the available data and the
medium scale variability timescale which we have chosen to explore.

However, the HEGRA points are not spread as uniformly over time as
are the BATSE and RXTE data. The ``center of gravity in observation time'' 
(defined as the weighted mean of the observation time of the daily points 
each weighted by the duration of the individual observation)
from HEGRA would therefore in general not coincide with that from BATSE 
and RXTE.
In order to compensate for this,  we calculate the center of gravity
in observation-time for each weekly HEGRA timebin and use these as
time-bin-{\it centers} for the other data. The edges of these bins
are then defined by the average of two adjacent bin-centers. 

Figure \ref{fig-lightcurve} shows the lightcurves  
from each instrument which went into this analysis.
The weekly points after the binning are given in 
table \ref{tab-data}. The first column in the table gives the
bin-centers as described above.
The following subsections describe the data used to compile the table.

\subsection{Optical and Radio Observations}

The optical observations were performed using the telescopes and 
filters listed in table~\ref{tab-instruments}. All the observations 
were made with CCD-cameras. All CCD-images were treated the normal way 
with flat field and bias corrections. The magnitudes were measured 
using either DAOPHOT in IRAF (NOT and Tuorla data; for more details 
see Katajainen  et al. \cite{katajainen}), or with the ROBIN-procedure 
developed in Torino (e.g. Villata et al. \cite{villata97}), or using
the automated reduction routine developed in Perugia (e.g. Tosti et al. 
\cite{tosti}). All magnitudes were measured using 10 arcsecond aperture. 
The use of the same aperture is important because Mkn~501 has a large 
host galaxy (e.g. Nilsson et al. \cite{nilsson}). For the calibration 
we used the calibration sequence given by Fiorucci \& Tosti 
(\cite{fiorucci}) and Villata et al. (\cite{villata98}). 

For some of the weekly observing periods, no optical data were 
available. However, the optical flux is extremely important for 
constraining the spectral index of the synchrotron spectra in 
the optical -- X-ray range, which, in turn, is essential for 
constraining the spectral index of the electron spectrum. As 
can be seen in table \ref{tab-data}, the optical flux exhibits 
only moderate variability on the time scales considered here. 
Thus, for fitting purposes we  assume that the optical flux 
during those viewing periods for which no optical observations 
were available, was within the range of optical fluxes observed 
during the whole campaign which yields $\nu F_{\nu} (opt.) = 
(7.485 \pm 0.774) \cdot 10^{-11}$~erg~cm$^{-2}$~s$^{-1}$.

The radio observations were made with the Mets\"ahovi Radio Telescope at 22 GHz
as part of the ongoing quasar monitoring program, which started in 1980.
Currently about 85 sources, mainly Northern flat spectrum quasars, should 
be observed monthly at 22 and 37 GHz. The total number of observations is 
now over 40000. For details of the observing strategy and reductions, as 
well as the data until 1995.5 see Ter\"asranta et al. (\cite{terasranta}).

The integration time of each point was essentially the same, so single points
were formed from the data points in the individual timebins
by simple averaging.

\subsection{Soft X-Ray Data}

Since the beginning of 1996, the all-sky monitor (ASM) of the
Rossi X-ray Timing Explorer (RXTE) satellite has been observing
Mkn 501 in the 2-12 keV energy band. The data used in this
analysis are taken from the publicly available ASM data products
provided by the ASM/RXTE teams at MIT and at the RXTE SOF and GOF 
at NASA's GSFC. These measurements are given by the authors (see 
e.g. Levine et al. \cite{levine}) as rates $R_1, R_2, R_3$ in units 
of counts per second in three energy bins: 1.3-3.0 keV, 3.0-5.0 keV, 
and 5.0-12.1 keV. From these we calculate the total rate $R$ by 
summing the three bins.

For Mkn 501 we obtain by averaging over the period MJD 50510 - 50710
\begin{equation}
  R_{mkn501} = 1.3 \pm 0.4
\end{equation}

In order to assess the X-ray spectrum and flux of Mkn 501, we use the 
ASM data for the Crab Nebula which is publicly available from the same 
source. The average rate of the Crab Nebula during the period MJD 50510 
-- 50710 is
\begin{equation}
   R_{crab} = 75.7 \pm 0.5\\
\end{equation}
Hence, in soft X-rays, Mkn 501 is even during this high state a significantly weaker 
source than the Crab Nebula.

In the energy range 2 keV to 60 keV, the spectrum of the Crab Nebula is a stable 
power law. The spectral index of the differential photon spectrum 
is known with an accuracy of 1\% to be
$\alpha_{crab} = 2.1$ (e.g. Toor \& Seward \cite{toor}, Pravdo \& Serlemitsos \cite{pravdo}
or Pravdo, Angelini \& Harding \cite{pravdo-b}).
The emission measured by the RXTE ASM is the sum of the steady Nebula and the pulsed
Crab Pulsar emission. The latter has on average a harder spectrum than the nebula.
Below 12.1 keV, however, the pulsed fraction of this emission is $< 8$~\%. 
We can therefore use the knowledge of the Crab Nebula spectrum to derive an approximate 
relative calibration factor $k$ for the rates $R_{2}$ and $R_{3}$ using
\begin{equation}
\frac{k R_{2,crab}}{R_{3,crab}} = \frac{E_3^{1-\alpha_{crab}} - E_2^{1-\alpha_{crab}}}
    {E_4^{1-\alpha_{crab}} - E_3^{1-\alpha_{crab}}} = 1.213  
\end{equation}
where $E_2 = 3.0$~keV, $E_3 = 5.0$~keV, and $E_4 = 12.1$~keV are the energy bin
edges of the second and third energy bin. We ignore the first energy bin since it
is strongly influenced by interstellar X-ray absorption which is dependent on the
column density and hence varies between sources (Remillard \cite{remillard}).

\begin{figure}
\epsfxsize=14.5cm
\epsffile{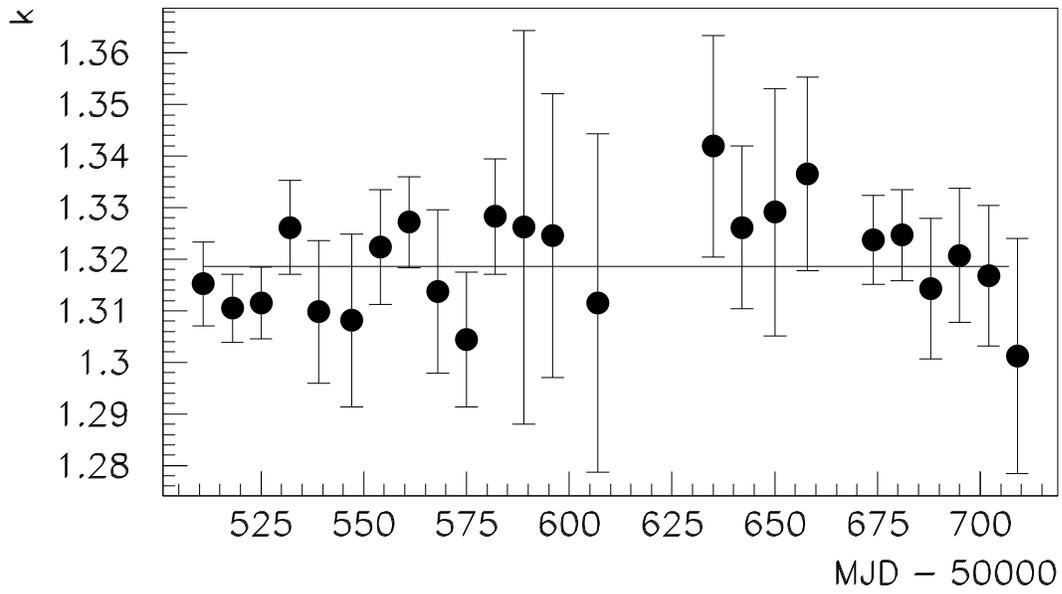}
\figcaption{\label{fig-k} 
The correction factor $k$ for the second RXTE ASM energy bin
derived from the publicly available Crab Nebula data
taken by the detector between MJD 50510 and 50710. See text.}
\end{figure}

Averaging over the available Crab data from the observation period 
under discussion (MJD 50510 - 50710) we obtain  $k = 1.318 \pm 0.0024$.
There is no indication of a variability of the value of $k$ (see figure \ref{fig-k}).
Hence we can assume that it is also valid for the Mkn 501 observations of the
same detector.
In order take into account that there is a pulsed component with a harder spectrum,
we add an additional error of 8 \% to the error of $R_3$ and obtain thus
\begin{equation}
 k = 1.32 \pm 0.023
\end{equation}

\begin{figure}
\epsfxsize=14.5cm
\epsffile{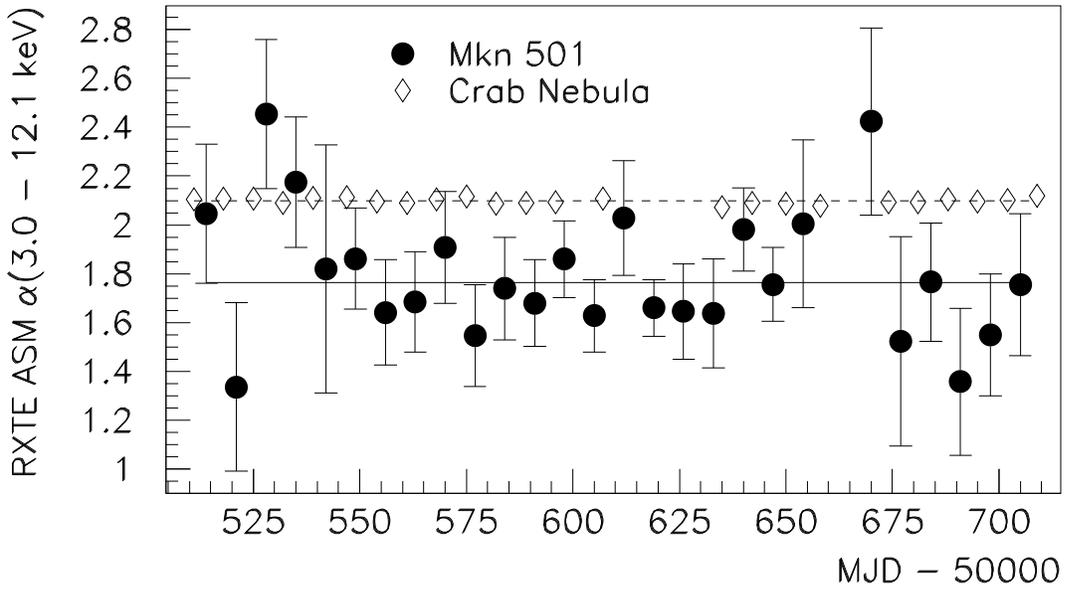}
\figcaption{\label{fig-alphas} 
The index $\alpha$ of the differential photon spectrum in the
energy range 3.0 - 12.1 keV derived from the publicly available
RXTE ASM data for Mkn 501 and the Crab Nebula in weekly time bins. 
The lines represent
fits of constant functions. For Mkn 501, the reduced $\chi^2$ of the fit is
0.93, for the Crab it is 0.95 . 
}
\end{figure}

For Mkn 501 we calculate the spectral index  $\alpha$ of the differential photon spectrum 
using the equation
\begin{equation}
\frac{k R_{2,mkn501}}{R_{3,mkn501}} = \frac{E_3^{1-\alpha} - E_2^{1-\alpha}}{E_4^{1-\alpha} - E_3^{1-\alpha}}  
\end{equation}
and varying $\alpha$ until the two sides of the equation are equal to an accuracy better
than 0.1~\%.
The error $\delta\alpha$ of this index, we estimate from the approximate formula
\begin{equation}
\alpha = -\frac{\log(k \cdot R_2/(5.0 - 3.0)) - \log(R_3/(12.1 - 5.0))}{\log((3.0 + 5.0)/(5.0 + 12.1))}
\end{equation}
which leads to
\begin{equation}
\delta\alpha = \sqrt{ 1.733 \cdot ( (\frac{\delta k}{k})^2 + (\frac{\delta R_2}{R_2})^2 + (\frac{\delta R_3}{R_3})^2 ) }
\end{equation}
where $\delta k$, $\delta R_2$, and $\delta R_3$ are the errors of the corresponding quantities.
The time resolved values for $\alpha_{mkn501}$ are shown in figure \ref{fig-alphas}.
The same method applied to the Crab Nebula data (however with $\delta k = 0.0024$) yields, 
as expected, on average the spectral index we have put in. Also this is shown in figure \ref{fig-alphas}.
From a constant fit to the values in this figure, we obtain
\begin{equation}
  \alpha_{mkn501} = 1.76 \pm 0.04
\end{equation}
and
\begin{equation}
   \alpha_{crab} = 2.10 \pm 0.02
\end{equation}
Thus we find that during the 1997 high state, Mkn 501 had on average a significantly harder 
soft X-ray spectrum than the Crab Nebula. 
The spectral variability of Mkn 501 is below the ASM's sensitivity.
The distribution of the weekly values is consistent with a constant value
(reduced $\chi^2$ = 0.93) and so is that for the Crab Nebula (reduced $\chi^2$ = 0.95).  

In order to calculate the $\nu F_{\nu}$ values in erg cm$^{-2}$ s$^{-1}$, 
we use the knowledge of the flux of the Crab Nebula. Here we face the problem that
there is still a disagreement of up to 25 \% between the measured normalization constants of the
Crab Spectrum from different experiments although there is perfect agreement in the spectral index. 
The discrepancy seems to stem from not very well understood 
systematic differences between the detectors (see the discussion in 
Pravdo, Angelini \& Harding \cite{pravdo-b}). 
We use the two most extreme recent measurements of the differential photon flux of the Crab Nebula together 
with the Crab Pulsar (pulse-averaged) and take their average as our normalization and their difference
as the systematic error of this quantity. From Pravdo \& Serlemitsos (\cite{pravdo})
we get at 5.2 keV a flux of (0.236 $\pm$ 0.006) photons~cm$^{-2}$s$^{-1}$keV$^{-1}$, while from
Pravdo, Angelini \& Harding (\cite{pravdo-b}) we get (0.302 $\pm$ 0.001) photons~cm$^{-2}$s$^{-1}$keV$^{-1}$.
The average of these values corresponds to a differential energy flux of 
\begin{equation}
F_{crab}(5.2 \mathrm{keV}) = (2.24 \pm 0.27_{syst}) \times 10^{-9} \mathrm{erg}\,\mathrm{cm}^{-2}\mathrm{s}^{-1}\mathrm{keV}^{-1}
\end{equation}
where the systematic error is the difference between the averaged values divided by 2.
The statistical errors of the two measurements are negligible. 
The energy flux of any other source with ASM rates $R_2$ and $R_3$ and differential photon spectral index
$\alpha$ is then calculated by
\begin{equation}
\begin{array}{rcl}
  \nu F_{\nu} [\mathrm{erg\,cm}^{-2} \mathrm{s}^{-1}] 
&  = &   (R_2 + R_3)[\mathrm{s}^{-1}]  \cdot  
  \frac{\displaystyle F_{crab}(5.2 \mathrm{keV})[\mathrm{erg}\,\mathrm{cm}^{-2}\mathrm{s}^{-1}\mathrm{keV}^{-1}]}
       {\displaystyle (R_{2,crab}+R_{3,crab})[\mathrm{s}^{-1}]} \\
& &  \cdot 5.2^{1 + \alpha_{crab} - \alpha} 
          \cdot \frac{\displaystyle 1 - \alpha}{\displaystyle 1 - \alpha_{crab}}
  \cdot \frac{\displaystyle E_4^{1 - \alpha_{crab}} - E_2^{1 - \alpha_{crab}}}
         {\displaystyle E_4^{1 - \alpha} - E_2^{2 - \alpha}} \\
& = & R[\mathrm{s}^{-1}] \cdot \frac{\displaystyle 5.2^{1 - \alpha} \cdot (1 - \alpha)}
             {\displaystyle 12.1^{1 - \alpha} - 3.0^{1 -\alpha}} \cdot 3.13 \times 10^{-10} \\
\end{array}
\end{equation}
where we correct for the difference in the spectra of Crab Nebula and the source in question
by making the Ansatz that the measured differential rates have the same ratio
as the differential fluxes. Furthermore we have used the result $R_{2,crab}+R_{3,crab} = 48.5 \pm 0.7$ s$^{-1}$
obtained from the dataset under discussion. 
Inserting the spectral index of Mkn 501  gives:
\begin{equation}
 \nu F_{\nu}(\mathrm{Mkn 501}, 5.2 \mathrm{keV})  =  
   (R_2 + R_3)[\mathrm{s}^{-1}] \cdot 2.40 \times 10^{-10} \mathrm{erg\,cm}^{-2} \mathrm{s}^{-1}
\end{equation} 
The error of this flux value is determined by propagating all errors of the quantities
involved which gives
\begin{equation}
\begin{array}{lr}
     \delta(\nu F_\nu)(\mathrm{Mkn 501}, 5.2 \mathrm{keV})  =  \\
   (\sqrt{5.76 \times 10^{-20} \cdot ((\delta R_2)^2 + (\delta R_3)^2) + 
          1.21 \times 10^{-23} \cdot (R_2 + R_3)^2}\,+\,0.12 \cdot \nu F_\nu)\, \mathrm{erg\,cm}^{-2} \mathrm{s}^{-1}
\end{array}
\end{equation}
where the term outside the square-root stems from the systematic error of the Crab flux normalization.

\subsection{Hard X-Ray Data}
\label{batse}

The Hard X-ray fluxes were measured using the Burst and Transient
Source Experiment  (BATSE)  onboard the CGRO satellite.
Although  BATSE is an uncollimated
detector, accurate point source fluxes can be measured using the
Earth Occultation method described in  
Harmon et al. (\cite{batsemethod}).  Daily sensitivities are 100
mCrab and over integrations of years, sources as weak as 3 mCrab can be
detected.  Several months are usually needed to obtain a statistically
significant flux from Mkn 501 with BATSE, but intense daily flares
can be seen, and the overall high state of 1997 allowed useful
measurements in each weekly interval even outside the flares.

The source is measured only when it sets and rises from
behind the Earth.
Since two such occultations occur per spacecraft orbit 
(roughly every 90 minutes), up
to 32  independent flux measurements can be made per day.
There exists, however,
a wide variation in this number
because of passage of the spacecraft through the South
Atlantic Anomaly, telemetry gaps from loss of TDRSS contact, and other
events, which occur randomly relative to the Mkn 501 steps.
For the data shown here between 61 and 211 measurements went into
an individual weekly point.  Each measurement lasts about 8
seconds giving a duty cycle of about 0.2\%.
In a 7 day time bin during the period discussed in this paper, BATSE
observes of the order of $10^5$ photons from the 
source (more during the flares).

The fluxes are integrated between 20 and 200 keV, and are
calculated by folding the measured counts through the BATSE detector response
assuming a differential source powerlaw spectrum of index -2.0.
The -2.0 spectrum is the best fit to the flare
 measured on MJD 50550-51 (between 20 and 1000 keV).
Spectra for other time intervals were also calculated  and were
consistent with -2.0. Uncertainties of 10~\% in the spectral index during 
flare times, larger
outside, make spectral variability difficult to assess for this source,
so that the index of -2.0 was used for each weekly interval.  Fluxes
were calculated for smaller
energy bands but the single 20-200 keV (median energy 36.4 keV) point 
for each interval is the most useful outside intense flares. 

Seven of the weekly averages are not statistically significant, and one period
shows a $1.8 \sigma$ deficit.  These eight points are inconsistent with 
a zero-level flux at the 99 \% confidence level.
The seven low but positive points are inconsistent with a zero-flux 
level at the 98-99 \% confidence level and are well fit by 
a constant flux of $1.1 \pm 0.3 \times
10^{-10}$ erg cm$^{-2}$ s$^{-1}$, implying that  
a flux is present which is below the sensitivity of the BATSE instrument in
weekly integrations.
An analysis of 30 blank fields on the sky shows that with 100
weekly integrations
for each field one can not distinguish (using an F-test) 
between a zero-flux level and the weighted mean of the weekly averages for
any of the 30 fields.  This implies that systematic effects are not 
responsible for the excess seen in the low points and that a flux is indeed
present.  For this reason these low but positive
values have been included as detections 
in the multiwavelength fits.   

\subsection{TeV data}

As described above, we use the data from the CT1 lightcurve of 
Mkn~501 (integral flux above 1.5 TeV) published by Aharonian et 
al. (\cite{hegrapartii}). These data are available in daily points 
based on observations of between 0.5~h and 5~h duration each. We 
group these points  according to our weekly time-bins and calculate
an average flux from the up to seven values weighting each daily point 
by its observation time.

Apart from giving daily flux measurements, the HEGRA papers
Aharonian et al. (\cite{hegraparti}), (\cite{hegrapartii}), and (\cite{hegraspec}) 
also determine the average spectral shape in the range between 0.5 and $\approx$ 25 TeV
with high accuracy. They find
\begin{equation}
\label{equ-spec}
dF/dE = N_0 E^{-\alpha} \exp(-E/E_0)
\end{equation}
where $N_0 = (10.8 \pm 0.2 \pm 2.1) \times
10^{-11}$~cm$^{-2}$~s$^{-1}$~TeV$^{-1}$, $\alpha = 1.92 
\pm 0.03 \pm 0.20$, and $E_0 = (6.2 \pm 0.4 \pm 2.2)$~TeV. 
The first error given is the statistical, the second the 
systematic error. Furthermore, they find that there is no 
spectral variability up to their sensitivity of $\delta\alpha 
\le 0.1$ on all relevant timescales. 

In order to include this important spectral information in our model fit,
we make the assumption that there is indeed no spectral variability and
extrapolate the points measured at 1.5 TeV by CT1 up to 10~TeV and down 
to 0.8~TeV. From the average integral flux values $F_{1.5}$ in units of 
photons cm$^{-2}$s$^{-1}$ we obtain the $\nu F_{\nu}$ values at photon 
energy $E$ in TeV using
\begin{equation}
\nu F_{\nu}(E) [\mathrm{erg \> cm}^{-2}\mathrm{s}^{-1}] = 1.6022 \cdot 
F_{1.5} \, {E^{(2 - \alpha)} \cdot \exp(-E/E_0) \over
\int\limits_{1.5}^{\infty} d\epsilon \> \epsilon^{-\alpha} \, 
\exp(-\epsilon/E_0)}.
\end{equation}
The extrapolation uses the measured spectral shape (equation 
\ref{equ-spec}) and fully propagates all statistical errors (error of the 
CT1 point, error of the spectral index $\alpha$ and the error of the 
cut-off energy $E_0$) to form the error of the extrapolated point.
The systematic errors of the spectral shape are expected to influence
all measured TeV spectra in the same way since they are caused by the 15~\%
uncertainty in the absolute energy calibration of the Cherenkov telescopes. 
Thus, for the purpose of a
spectral variability study, the statistical errors are those which
actually determine the uncertainty in the differences of the spectral
shape between different measurements. For this reason we propagate
only the statistical errors in the extrapolation. In this way, the 
columns 6 and 8 of table \ref{tab-data} are obtained. In the model fit 
(section \ref{sec-fitting}) we take into account the uncertainty in the 
energy scale by introducing errors of $\pm 15$~\% along the energy axis.

The energies to which we extrapolate are chosen as a compromise of maximum
distance to 1.5 TeV and minimum systematic errors. The latter are
increasing up to several 10\,\% towards both ends of the range for 
which the spectrum has been measured (see Aharonian et al. 
\cite{hegraspec}, figure 9) but are still small at 0.8 and 10 TeV.

The correlation between the three TeV points which we introduce into
the model fit by performing the described extrapolation is not problematic
since we do not plan to calculate absolute $\chi^2$ values in the fits for
proving that the model describes the data better than another.
Instead, the fits serve the aim to study the time-dependent behaviour of 
the model parameters.
The extrapolated points are only a means to take into account the available
spectral information.

\subsection{Fitting the model to the SEDs}
\label{sec-fitting}

To each weekly SED we fit the Blazar jet model described in detail
by B\"ottcher et al. (\cite{boettcher97}). The model assumes that
isolated components (blobs) of relativistic pair plasma, which
are assumed to be spherical in the co-moving frame, are 
injected instantaneously into the jet, and follows the 
self-consistent evolution of the particle and radiation spectra 
as the blob moves outward along the jet, taking into account
all relevant radiation, cooling, and absorption mechanisms: 
synchrotron radiation, synchrotron self-absorption, synchrotron
self-Compton scattering, external Compton scattering of direct 
accretion disk radiation, $\gamma$-$\gamma$ absorption and pair 
production intrinsic to the source. The magnetic field is
assumed to decay along the jet as $B \propto r^{-1}$, 
where $r$ is the distance from the center of the AGN. 
The emerging, time-averaged spectra are corrected for 
$\gamma\gamma$ absorption by the intergalactic infrared 
background radiation using the lower model spectrum given 
by Malkan \& Stecker (\cite{malkan98}). 

Since we are interested in weekly averages, the emission from 
single blobs is time-integrated over the jet evolution and
subsequently re-converted into a flux by dividing the fluence
by an average repetition time $\Delta t_{\rm rep}$ of blob 
ejection events. We assume that a fraction $f < 1$ of the jet 
is filled with relativistic pair plasma. The filling factor 
is given by $f \sim R'_B / (\Gamma \, c \, \Delta t_{\rm rep})$,
where $R'_B$ is the blob radius in the co-moving frame and
$\Gamma$ is the bulk Lorentz factor of the blobs. Even if 
the filling factor is close to unity, it is a reasonable 
approximation to assume that the blobs do not interact 
with each other because the synchrotron and SSC radiation 
produced within the jet are isotropic in the comoving 
frame so that most of the radiation escapes to the sides 
without interaction with the rest of the jet.

As mentioned in the introduction, extreme HBLs like Mkn~501 
or Mkn~421 are generally well described by a pure SSC model. 
A simple analytic estimate shows that the observed TeV 
$\gamma$-ray spectrum can not plausibly be produced by 
Comptonization of radiation from an accretion disk around 
the putative supermassive black hole in the center of 
Mrk~501: The spectrum emitted by an optically thick, 
geometrically thin accretion disk is reasonably well
approximated by a blackbody spectrum whose temperature, for
a black hole mass of $\gtrsim 10^8 \, M_{\odot}$, yields an
average photon energy of the disk radiation of $\epsilon_D
\equiv h \nu_D / (m_e c^2) \sim 10^{-5}$. If external Comptonization
is to be efficient in competition with the synchrotron self-Compton
process, the blob has to be rather close to the accretion disk,
or a significant fraction of the disk photons has to be rescattered 
into the jet by surrounding material, so that the bulk of disk 
photons enters the blob from the side and is blue shifted 
by a factor of $\sim \Gamma$ into the blob rest frame. 
Thus, due to the strong reduction of the Klein-Nishina cross 
section for $\gamma_e \epsilon' \gtrsim 1$ (where $\epsilon'$
is the photon energy in the comoving frame), no significant 
radiative output at (observer's frame) energies $\epsilon_{obs} 
\gtrsim D / (\epsilon_D \Gamma) \sim 10^5$ (where $D$ is the 
Doppler factor), corresponding to $E_{obs} \gtrsim 100$~GeV, 
will be produced by external Comptonization, ruling out this 
process to explain the observed high-energy spectrum extending 
to TeV energies. If the high-energy spectrum in the $\gtrsim 
100$~GeV regime is produced by the SSC process, then the level 
of SSC radiation at 1~GeV may be estimated by
\begin{equation}
\nu F_{\nu}^{SSC} (1 \, {\rm GeV}) \approx \nu F_{\nu}^{SSC}
(E_{pk}) \, \left( {E_{pk} \over 1 \, {\rm GeV}} \right)^{p - 3
\over 2}
\label{GeV_estimate}
\end{equation}
where $E_{pk} \gtrsim 0.1$~TeV is the energy of the $\nu F_{\nu}$ 
peak in the high-energy part of the spectrum and $p$ is the spectral 
index of the injected electron spectrum. Inserting a typical value 
of $\nu F_{\nu} (E_{pk}) \gtrsim 10^{13} \> {\rm Jy \, Hz}$ and 
$p \sim 2.5$, this yields $\nu F_{\nu}^{SSC} (1 \, {\rm GeV})
\gtrsim 3 \cdot 10^{12} \> {\rm Jy \, Hz}$, which is already close
to the maximum ever observed of $\sim 10^{13} \> {\rm Jy \, Hz}$.
In reality, the SSC spectrum is not a straight power-law below
$E_{pk}$, but shows a gradual turnover so that the actual SSC
flux at 1~GeV is substantially higher than the above estimate, 
leaving little room for an additional EIC contribution. For our
fitting procedure, we are thus using a simplified version of
our jet simulation code, in which the photon output (but not
the electron cooling) from Comptonization of accretion disk
radiation is neglected. These simulations properly account for 
the self-consistent cooling of the electron population and 
$\gamma\gamma$ absorption and pair production in the blob. 
In the simulations, the exact, angle-averaged synchrotron 
spectrum of an isotropic electron population as given by 
Crusius \& Schlickeiser (\cite{crusius86}) is used.

\begin{figure}
\epsfysize=15cm
\rotatebox{270}{\epsffile{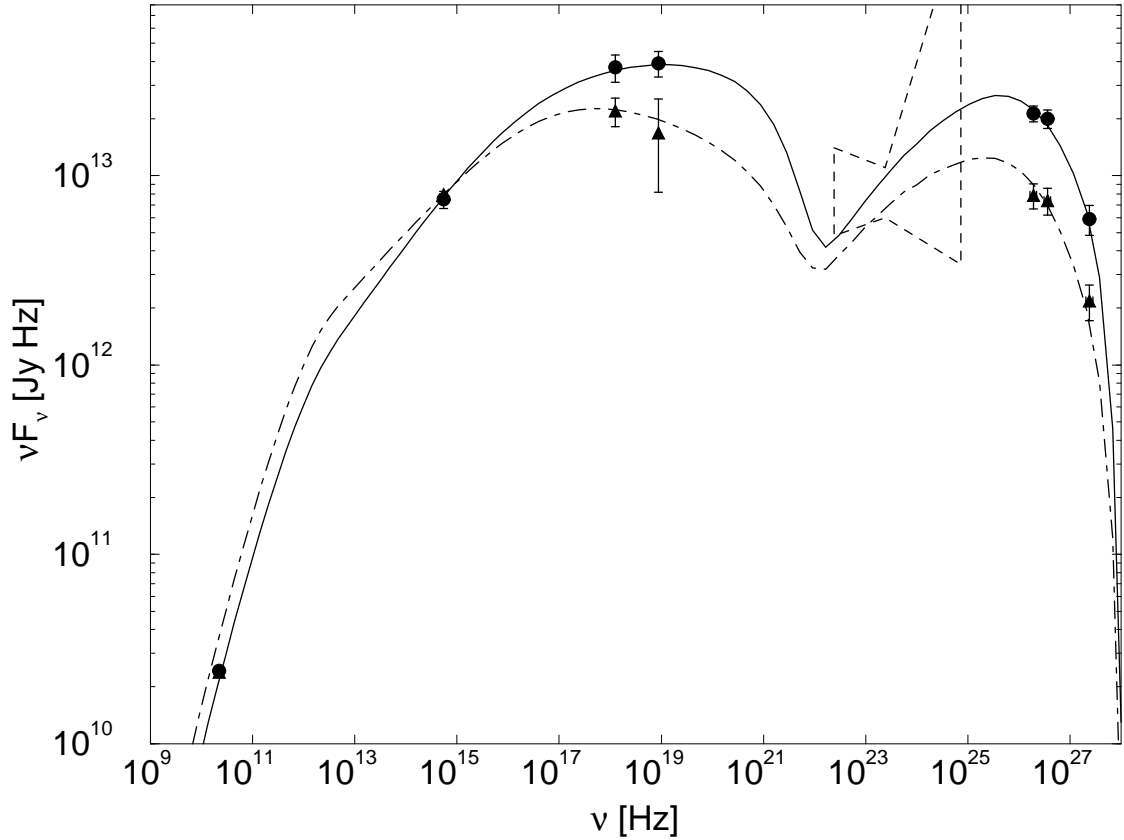}}
\figcaption{\label{ssc_fits}
Model fits to the weekly averaged broadband
spectra of Mkn~501 for the periods centered on 
MJD~50564.566 (low flux state; filled triangles
and dot-dashed curve) and MJD~50626.785 (high 
flux state; filled circles and solid curve). Model 
parameters for MJD50564.566: $\gamma_1 = 500$, 
$\gamma_2 = 3 \cdot 10^7$, $p = 2.400$, $n_e = 
15$~cm$^{-3}$, $\Delta t_{rep} = 1.74 \cdot 10^3$~s, 
$B = 0.05$~G, $\Gamma = 25$, $R'_B = 3 \cdot 10^{15}$~cm, 
$D = 30$. Model parameters for MJD~50626.785: Same as 
for the low state, except $p = 2.225$, $\Delta t_{rep} 
= 3.0 \cdot 10^3$~s.} 
\end{figure}

Fig. \ref{ssc_fits} shows two examples of fits to low 
and high flux levels of Mkn~501. The respective fit 
parameters are given in the figure captions. Inverse-Compton 
cooling on accretion-disk photons may be important close to 
the base of the jet even if the resulting photon spectra do
not contribute significantly to the time-averaged emission
and is thus still included in our simulations.

Due to the non-linear nature of the model system, the fit results
cannot be described in a simple, linear way as a function of the
model parameters. We therefore construct a three-dimensional mesh
of simulations in parameter space, with the electron density $n_e$, 
the high-energy cutoff $\gamma_2$ and the spectral index $p$ of the 
injected electron spectrum  as parameters which are free to vary on 
the grid points. 

We calculate our parameter grid varying $p$ in steps of 0.025 between
1.6 and 2.8, the total electron number density $n_e$ in steps
of 5~cm$^{-3}$ between 10 and 120~cm$^{-3}$, and $\gamma_2$
for values of $2 \cdot 10^6$, $3 \cdot 10^6$, $5 \cdot 10^6$,
$7.5 \cdot 10^6$, $10^7$, $1.5 \cdot 10^7$, $2 \cdot 10^7$, 
$2.5 \cdot 10^7$, and $3 \cdot 10^7$. We constrain the range 
of $\gamma_2$ values to $\gamma_2 \le 3 \cdot 10^7$ because of 
the kinematic limit and because around this energy, electron 
cooling due to triplet pair production on the highest-energy 
synchrotron photons, which is ignored in our simulations, becomes 
dominant over Compton scattering (Mastichiadis et al. 
\cite{mastichiadis94}, Anguelov et al. \cite{ang99}). 
We find that our simulated spectra are only very weakly 
dependent on the actual value of $\gamma_2$. A change of 
$\gamma_2$ by a factor 3 typically results in an increase 
of the reduced $\chi^2$ of only 0.3 such that the 
above-mentioned restriction of $\gamma_2$ values has
only minor impact on our results. In fact $\gamma_2$ can
be regarded as constant and of the order of $10^7$. 
We point out that the instantaneous synchrotron spectra of
individual blobs at the time of injection, which might
correspond to short-term X-ray and TeV flares, have 
their synchrotron peak at $\nu_{sy, inst.} \sim 2.8 \cdot
10^6 \> (B/{\rm G}) \, D \, \gamma_2^2$~Hz if $p < 3$, in 
agreement with the shift of the synchrotron peak into the 
hard X-ray regime during extreme flaring activity (e. g., 
Pian et al. \cite{pian}).

All other parameters (in particular, the magnetic field at
the particle injection site, $B_0 = 0.05$~G, the low-energy
cutoff of the electron spectrum $\gamma_1 = 300$, and the 
Doppler factor, $D = 30$) are fixed to values allowing good 
fits to the observed weekly averaged SEDs using our simulation 
code. An estimate for the required parameters can be found
on the basis of the location of the synchrotron and SSC peaks
of the observed broadband spectra as described below.

Although we are assuming the injection of a single power-law
distribution of ultrarelativistic electrons into the jet, the 
time-averaged radiation spectrum will be reasonably well 
approximated by the one produced by a broken power-law 
distribution of electrons with spectral index $p$ below 
the break energy $\gamma_b$, and $p + 1$ above the break 
energy. $\gamma_b$ may be computed by setting the synchrotron 
cooling time scale equal to the dynamical time scale of jet 
evolution (magnetic field decay), which yields
\begin{equation}
\gamma_b \approx 6.4 \cdot 10^5 \> {\Gamma_{25} \over
z_{0.03} \, B_{-1}^2}
\label{gamma_b}
\end{equation}
where $\Gamma_{25} \equiv \Gamma / 25$, the height of the
injection/acceleration site above the accretion disk is 
$z_i = 0.03 \, z_{0.03}$~pc, and $B_{-1} = B_0 / (0.1 \, 
{\rm G})$. Thus, our model calculations will produce a 
time-averaged synchrotron break at
\begin{equation}
\nu_{sy} \sim 3.4 \cdot 10^{18} \> \overline B_{-1} \, D_{30}
\, {\Gamma_{25}^2 \over z_{0.03}^2 \, B_{-1}^4} \> {\rm Hz}
\label{nu_sy}
\end{equation}
where $\overline B_{-1}$ is an appropriate average of the 
magnetic field (in units of $0.1$~G) over the jet evolution. 
For the purpose of these estimates, we neglect factors of
$(1 + z) \sim 1$ for Mkn~501. The location of the peak of 
the SSC component will be strongly influenced by Klein-Nishina 
effects and will thus depend on the actual shape of the 
synchrotron spectrum, which, in turn, depends on the electron 
spectral index $p$. Considering these effects, Tavecchio et 
al. (\cite{tav98}) find
\begin{equation}
\epsilon_{SSC} \sim \gamma_b \, D \, g(\alpha_1, \alpha_2)
\label{epsilon_ssc}
\end{equation}
where $\alpha_1 = (p - 1)/2$, $\alpha_2 = p/2$, and
\begin{equation}
g(\alpha_1, \alpha_2) = \exp\left[ {1 \over \alpha_1 - 1}
+ {1 \over 2 \, (\alpha_2 - \alpha_1)} \right].
\label{g_alpha}
\end{equation}
For $p = 2.5$, this yields $\epsilon_{SSC} \sim 9.4 \cdot 10^5
\> D_{30} \, \Gamma_{25} / (z_{0.03} \, B_{-1}^2)$, corresponding
to
\begin{equation}
E_{SSC} \sim 490 \> { D_{30} \, \Gamma_{25} \over z_{0.03} \,
B_{-1}^2} \> {\rm GeV}.
\label{E_ssc}
\end{equation}
Combining Eqs. \ref{nu_sy} and \ref{E_ssc} and using the average
observed $\epsilon_{sy} \sim 10^{-2}$ and $\epsilon_{SSC} \sim 10^6$,
we find
\begin{equation}
{\overline B_{-1} \over D_{30}} \sim  300 {\epsilon_{sy} \over 
\epsilon_{ssc}^2} \, g^2 (\alpha_1, \alpha_2) \sim 0.34
\label{BD_1}
\end{equation}
for $p = 2.5$ (see also Tavecchio et al. \cite{tav98}). Similarly,
we may use Eq. (22) of Tavecchio et al. (\cite{tav98}) to estimate
\begin{equation}
\overline B \, D^{2 + \alpha_1} \gtrsim  \left[ g(\alpha_1, \alpha_2)
\, \epsilon_{SSC} \, \epsilon_{sy} \right]^{(1 - \alpha_1)/2} \, \sqrt{
2 f(\alpha_1, \alpha_2) \over c^3} \> {\left( \nu L_{\nu} \right)_{sy}
\over t_{var} \, \sqrt{ \left( \nu L_{\nu} \right)_{SSC}}}
\end{equation}
where $f(\alpha_1, \alpha_2) = 1/(1 - \alpha_1) + 1/(\alpha_2 - 1)$.
Using $p = 2.5$, $\left(\nu L_{\nu} \right)_{sy} \sim 6 \cdot 10^{43}$~erg/s,
$\left(\nu L_{\nu} \right)_{SSC} \sim 2 \cdot 10^{44}$~erg/s, and
$t_{var} \sim 5$~h (Aharonian et al. \cite{hegraparti}), we find
\begin{equation}
\overline B_{-1} \, D_{30}^{11/4} \gtrsim 0.34.
\label{BD_2}
\end{equation}
Combining this with Eq. \ref{BD_1}, we have $D_{30} \gtrsim 1$ and
$\overline B_{-1} \gtrsim 0.34$. These numbers are consistent with
the limits found by Bednarek \& Protheroe \cite{bednarek99}, but
are slightly outside the allowed region of parameter space as found
by Tavecchio et al. (\cite{tav98}) and Kataoka et al. (\cite{kataoka}).
This is because in those papers the broadband spectrum is either
characterized by quantities pertaining to individual outbursts or
to a long-term quiescent state. Those parameters are not representative
of the weekly averages investigated in this paper. 

Having constructed the three-dimensional mesh of simulations,
we compare all weekly SEDs with the simulated spectra and find
the simulation with the smallest $\chi^2$. Results of this
procedure are described in the next section.

\section{Results}

\subsection{Correlation TeV-X-Ray}

The SSC model for Mkn~501 predicts a very strong correlation 
between the emission at the synchrotron peak (in soft -- hard 
X-rays) and at the inverse-Compton peak (close to TeV 
$\gamma$-ray energies). We have derived an analytic estimate
for the expected correlation, for variations of several input
parameters. In the following discussion, unprimed quantities are
measured in the co-moving frame, while a superscript $\ast$ refers
to quantities measured in the observer's frame. We assume that 
the time-averaged (cooling) electron spectrum can be described 
by a broken power-law, 

\begin{equation}
n_e (\gamma) = n_0 \cases{ (\gamma / \gamma_b)^{-p} & 
for $\gamma_1 \le \gamma \le \gamma_b$ \cr
(\gamma / \gamma_b)^{-(1 + p)} 
& for $\gamma_b \le \gamma \le \gamma_2$ \cr}
\end{equation}
where the injection spectral index $2 < p < 3$, and $\gamma_b$ 
is the break energy of the spectrum, determined by Eq. \ref{gamma_b}. 
The normalization is given by $n_0 \approx n_e \, \gamma_b^{-p}
\, \gamma_1^{p-1} / (p - 1)$. We are using a $\delta$ approximation 
for the synchrotron spectrum:
\begin{equation}
L_{sy} (\epsilon) = L_0 \cdot \cases{ (\epsilon / \epsilon_b)^{1 - p 
\over 2} & for $\epsilon_1 \le \epsilon \le \epsilon_b$ \cr
(\epsilon / \epsilon_b)^{-{p \over 2}} 
& for $\epsilon_b \le \epsilon \le \epsilon_2$, \cr}
\end{equation}
where $\epsilon = h \nu / (m_e c^2)$ is the dimensionless photon energy
and $\epsilon_i = 2.3 \cdot 10^{-14} \, (B/ {\rm G}) \, \gamma_i^2$ 
is the characteristic synchrotron energy radiated by an electron of 
Lorentz factor $\gamma_i$. Normalizing the synchrotron luminosity to

\begin{equation}
L_{sy} \propto B^2 \int\limits_{\gamma_1}^{\gamma_2} d\gamma \>
n_e (\gamma) \, \gamma^2,
\label{Lsy}
\end{equation}
we have 
\begin{equation}
L_0 \propto {B \, n_e \over p - 1} \, \left( {\gamma_1 \over
\gamma_b} \right)^{p-1}.
\label{norm_L}
\end{equation}

Neglecting Compton scattering events in the Klein-Nishina
regime, $\epsilon\gamma > 3/4$, we may approximate the SSC 
spectrum by

\begin{equation}
L_{SSC} (\epsilon_s) \propto \int\limits_{\epsilon_1}^{\epsilon_2} 
d\epsilon \> {L_{sy} (\epsilon) \over \epsilon} \, \sqrt{\epsilon_s 
\over \epsilon} \, n_e \left( \sqrt{3 \, \epsilon_s \over 4 \, 
\epsilon} \right) \, \Theta\left( {3/4} - \sqrt{\epsilon_s \, 
\epsilon} \right),
\label{LSSC}
\end{equation}
where $\Theta$ is the Heaviside function. The evaluation of this
expressions is straightforward. Observed fluxes in the ASM, BATSE,
and HEGRA energy ranges are calculated integrating the Doppler
boosted synchrotron and SSC spectra, $L^{\ast} (\epsilon^{\ast})
= D^3 \, L (\epsilon^{\ast}/D)$, over the energy ranges 
$4 \cdot 10^{-3} \le \epsilon^{\ast}_{ASM} \le 2 \cdot 10^{-2}$,
$4 \cdot 10^{-2} \le \epsilon^{\ast}_{BATSE} \le 0.4$, and
$3 \cdot 10^6 \le \epsilon^{\ast}_{HEGRA} \le 6 \cdot 10^7$.

\begin{figure}\centering
\epsfysize=12cm
\rotatebox{270}{\epsffile{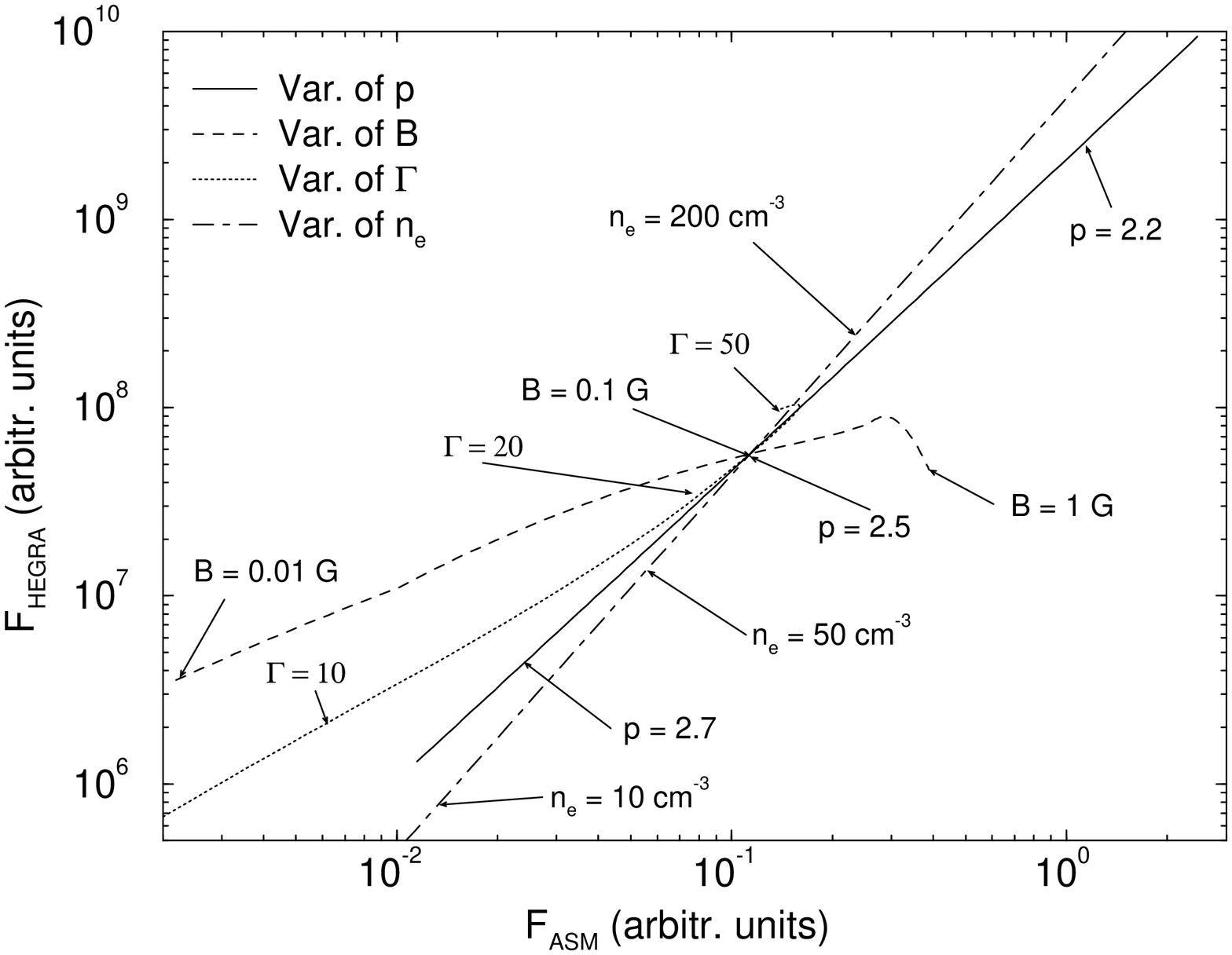}}
\vskip 1cm
\epsfysize=12cm
\rotatebox{270}{\epsffile{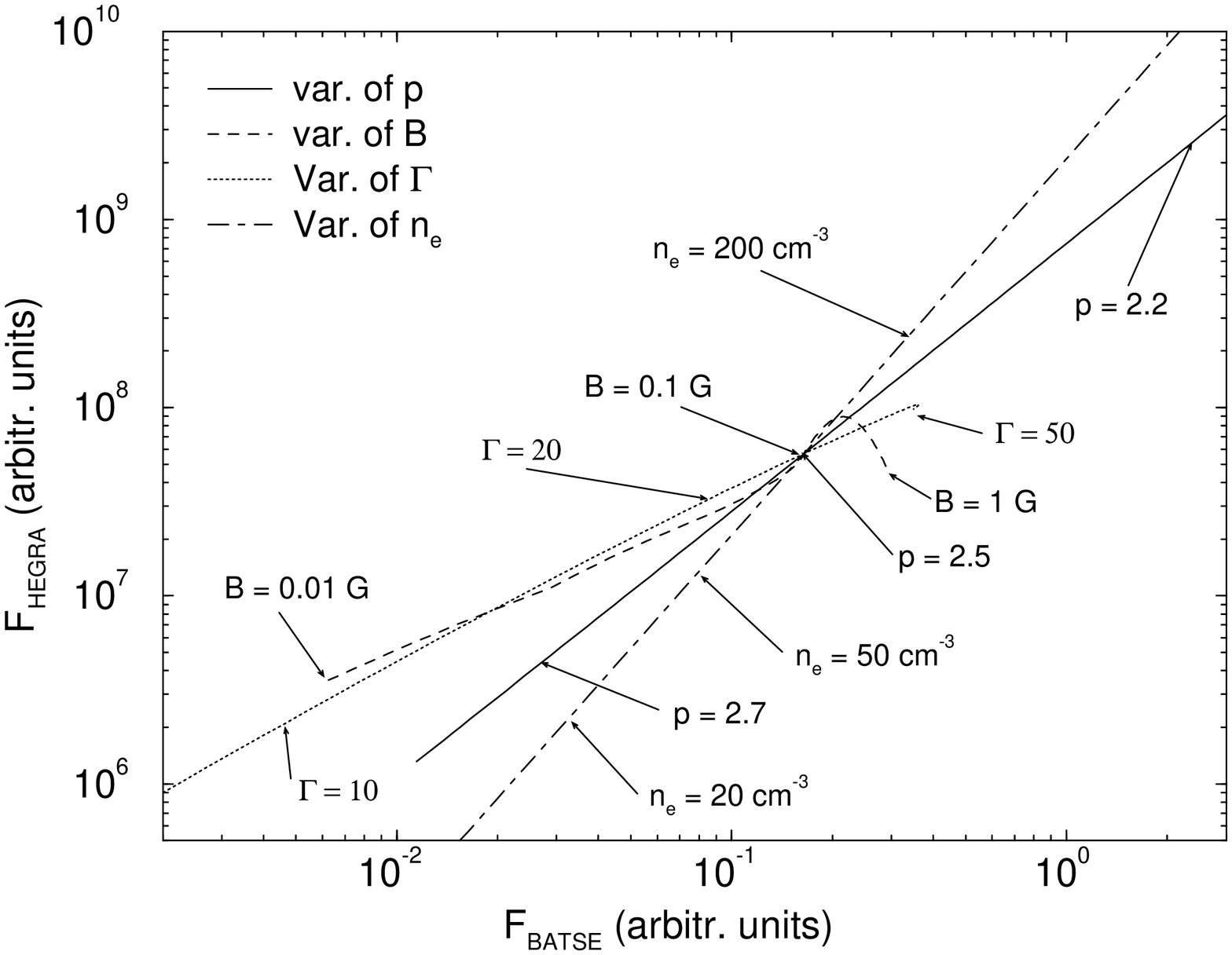}}
\figcaption{\label{fig_ssc_var}
Correlation between the ASM and HEGRA fluxes (a) and between
the BATSE and HEGRA fluxes (b) according to our analytical 
approximation. Standard model parameters are $\gamma_1 = 300$, 
$B = 0.1$~G, $n_e = 100$~cm$^{-3}$, $p = 2.5$, $\gamma_2 
= 10^7$. For each curve, one parameter is varied, while
the others are fixed to the above values. The curves are
labelled by a few representative values of the varying
parameter.}  
\end{figure}

In Fig. \ref{fig_ssc_var}, we plot trajectories in the $(F_{BATSE}, 
F_{HEGRA})$ and $(F_{ASM}, F_{HEGRA})$ planes of these solutions, 
varying individual model parameter separately while all others are 
held constant at values representative of states of moderate X-ray
and high-energy $\gamma$-ray fluxes.

A variation of the electron density obviously yields a
relation $F_{SSC} \propto F_{sy}^2$ since the synchrotron flux
depends linearly, the SSC flux quadratically on $n_e$. Note that
this dependence may be altered due to an increasing $\gamma\gamma$
absorption opacity intrinsic to the source, which is not included 
in the analytical estimate (\ref{LSSC}) used to compute the HEGRA 
flux. A variation of the electron injection spectral index $p$ 
results in a relation which may be approximated by $F_{HEGRA} 
\propto F_{BATSE}^{1.4}$ and $F_{HEGRA} \propto F_{ASM}^{1.6}$, 
i. e. the dependence is weaker than quadratic. 

A variation of the magnetic field strength leads to more complex 
flux variations due to the back-reaction on the break Lorentz
factor $\gamma_b$ as a result of radiative cooling. For relatively
strong magnetic fields ($B \gtrsim 0.3$~G), the X-ray and TeV
$\gamma$-ray fluxes become anti-correlated.

Finally, if the variability of this source were dominated by a
variation of the bulk Lorentz factor $\Gamma$, the X-ray and
high-energy $\gamma$-ray fluxes would be expected to be approximately
linearly correlated (the back-reaction on $\gamma_b$ leads to a
slight deviation from a strictly linear correlation), as long as
the observer is located within the $1/\Gamma$ beaming cone of
the jet. If $\Gamma$ increases beyond $\sim 1/\theta_{obs}$,
both the X-ray and TeV $\gamma$-ray fluxes start to decrease
with increasing $\Gamma$. The same quasi-linear correlation
would be expected if the variability were due to a bending
jet, i. e. a variation of $\theta_{obs}$.

\begin{figure}
\centering
\epsfxsize=14.5cm
\epsffile{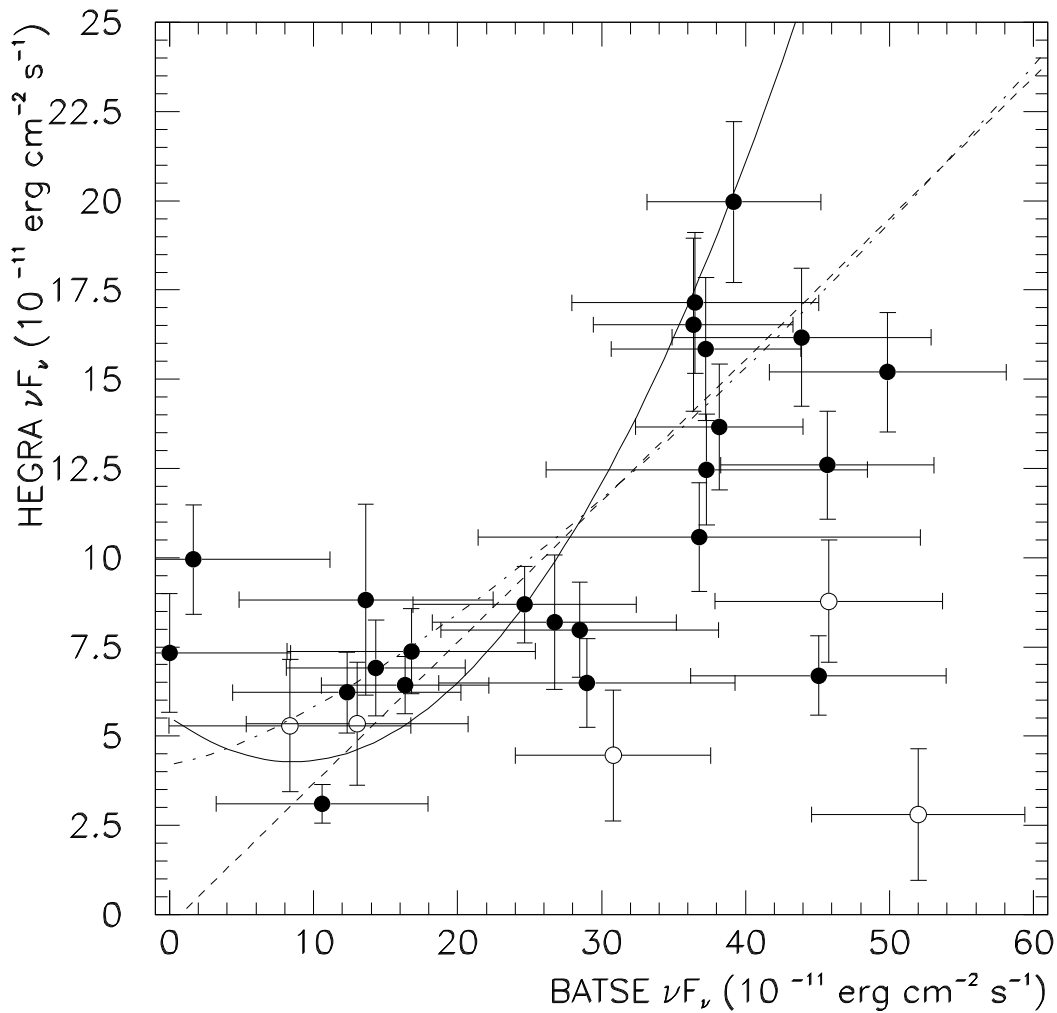}
\figcaption{\label{sy_ssc_corr}
The correlation between HEGRA TeV and the BATSE X-ray 
flux from Mkn~501. In order to avoid effects from
poor time-coverage, points obtained from less than three
independent HEGRA measurements were excluded. The excluded
points are shown as open circles. The fit of a second-order
polynomial to the remaining points (filled circles) yields 
$y = (5.4 \pm 0.6) + (- 0.29 \pm 0.08) x + (0.017 \pm 0.0035) x^2$ (solid line), 
where $y = $~HEGRA flux at 1.5 TeV in units of $10^{-11}$~erg~cm$^{-2}$~s$^{-1}$,
and $x = $~BATSE flux at 36.4 keV in the same units, and results in a reduced
$\chi^2$ of $1.14$. With the open circle points included the reduced 
$\chi^2$ increases to 1.54. The dashed line is a linear fit to the
filled circle points: $y = (-0.3 \pm 1.7) + (0.40 \pm 0.06) x$,
reduced  $\chi^2 = 1.42$ (2.6 with the open circle points).
The dot-dashed line is a fit of the function $y = a x^{1.4} + b$. It results
in $a = 0.063 \pm 0.0084$, $b = 4.2 \pm 0.72$, and a reduced $\chi^2$ of 1.65 
(2.7 with the open circle points).}
\end{figure}

The empirical correlation of the TeV and the X-ray emission of 
Mkn~501 in 1997 has already been studied extensively using the 
CT1/CT2 and the CTS data from HEGRA and the soft X-ray data from 
RXTE ASM: Aharonian et al. (\cite{hegrapartii}) find the correlation 
coefficient for the daily averages to be $0.61 \pm 0.057$. This 
maximum correlation is found for zero time-lag. We examine 
the correlation between the weekly RXTE, BATSE 
and HEGRA points from table \ref{tab-data}. Fig. \ref{sy_ssc_corr}
shows the observed correlation between the HEGRA and BATSE
measurements, fitted with a second-order
polynomial as well as with a power-law with
index 1.4, which is the theoretical prediction if the variability is 
caused solely by variations of the electron
injection spectral index $p$. Both fits give acceptable values for
the reduced $\chi^2$  (1.1 and 1.65 respectively). Still, due to 
the large error bars, we can not confidently distinguish between 
these and similar correlations on the basis of the currently 
available data. In any case, there is no indication for a 
super-quadratic dependence between the X-ray and TeV fluxes, 
which would be inconsistent with a pure SSC model, unless
there is a persistent quiescent level of emission above
which the observed flaring behaviour is superimposed.

\begin{figure}
\centering
\epsfxsize=14.5cm
\epsffile{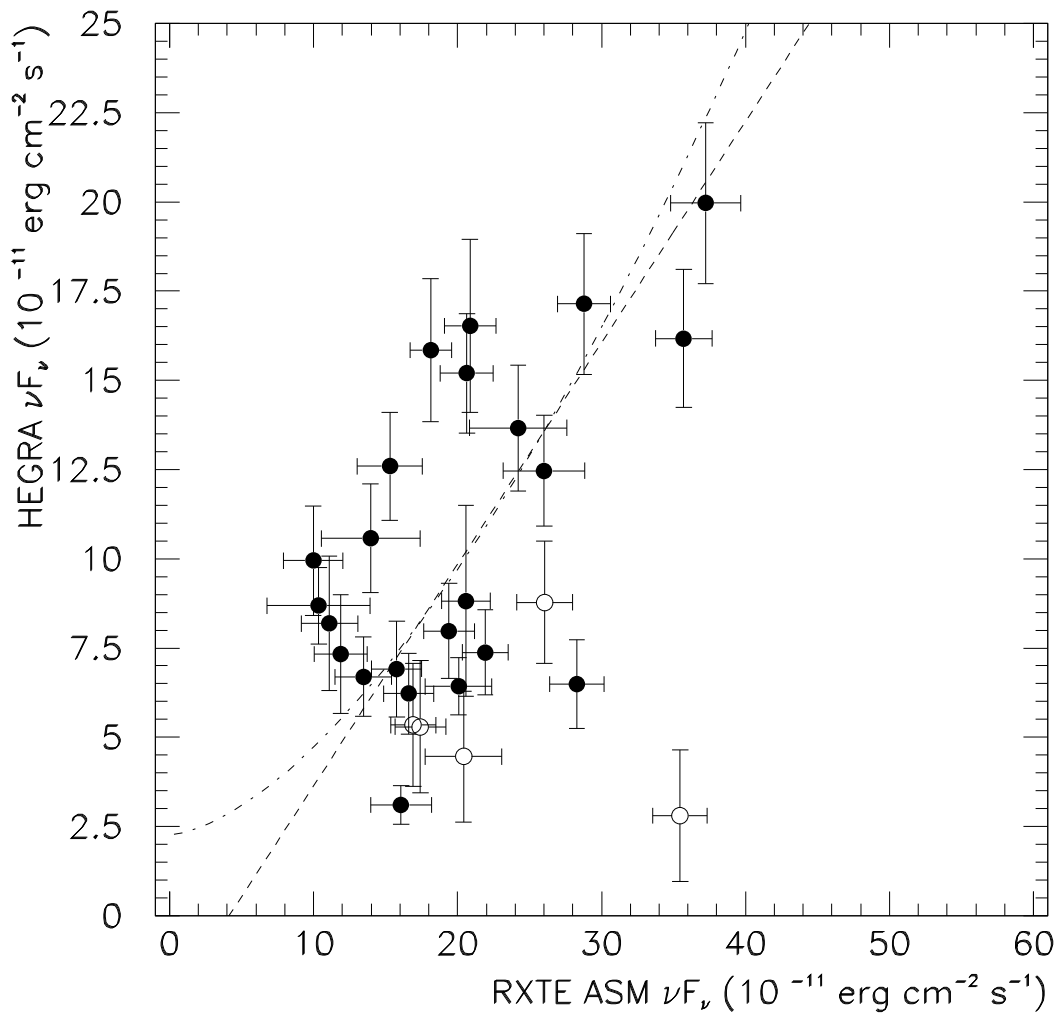}
\figcaption{\label{fig-hegra-rxte}
The correlation between the HEGRA TeV and the RXTE ASM X-ray 
flux from Mkn~501. In order to reduce effects from
poor time-coverage, points obtained from less than three
independent HEGRA measurements were excluded. The excluded
points are shown as open circles. 
The fit of a linear function to the remaining points (filled circles) yields 
$y = (-2.6 \pm 1.40) + (0.62 \pm 0.07) x$, where
$y = $~HEGRA flux at 1.5 TeV in units of $10^{-11}$~erg~cm$^{-2}$~s$^{-1}$,
and $x = $~RXTE flux at 5.2 keV in the same units, and results in a reduced
$\chi^2$ of $4.7$. Only statistical errors were taken into account.
The linear correlation coefficient is $0.59$. 
The dot-dashed line is a fit of the function $y = a x^{1.6} + b$ which results
in $a = 0.062 \pm 0.0075$, $b = 2.28 \pm 0.80$, and a reduced $\chi^2$ of 4.9 .}
\end{figure}

Figure \ref{fig-hegra-rxte} shows the correlation between the weekly
TeV and Soft-X-ray points. The linear correlation coefficient
is $0.59$, nearly the same as found by Aharonian et al. (\cite{hegrapartii})
who compare the same data on a daily basis.
The constant term of the linear fit is still consistent with zero.
However, the reduced  $\chi^2$ of 4.7 is too large for a good fit. 
This is also the case for a fitted power-law with index 1.6 which
gives $\chi^2 = 4.9$. The systematic differences in 
time-coverage which are not taken into account in the determination
of the error bars may be responsible for this.
Still, the large linear correlation coefficient suggests that the
correlation is nearly linear.  

\begin{figure}
\centering
\epsfxsize=14.5cm
\epsffile{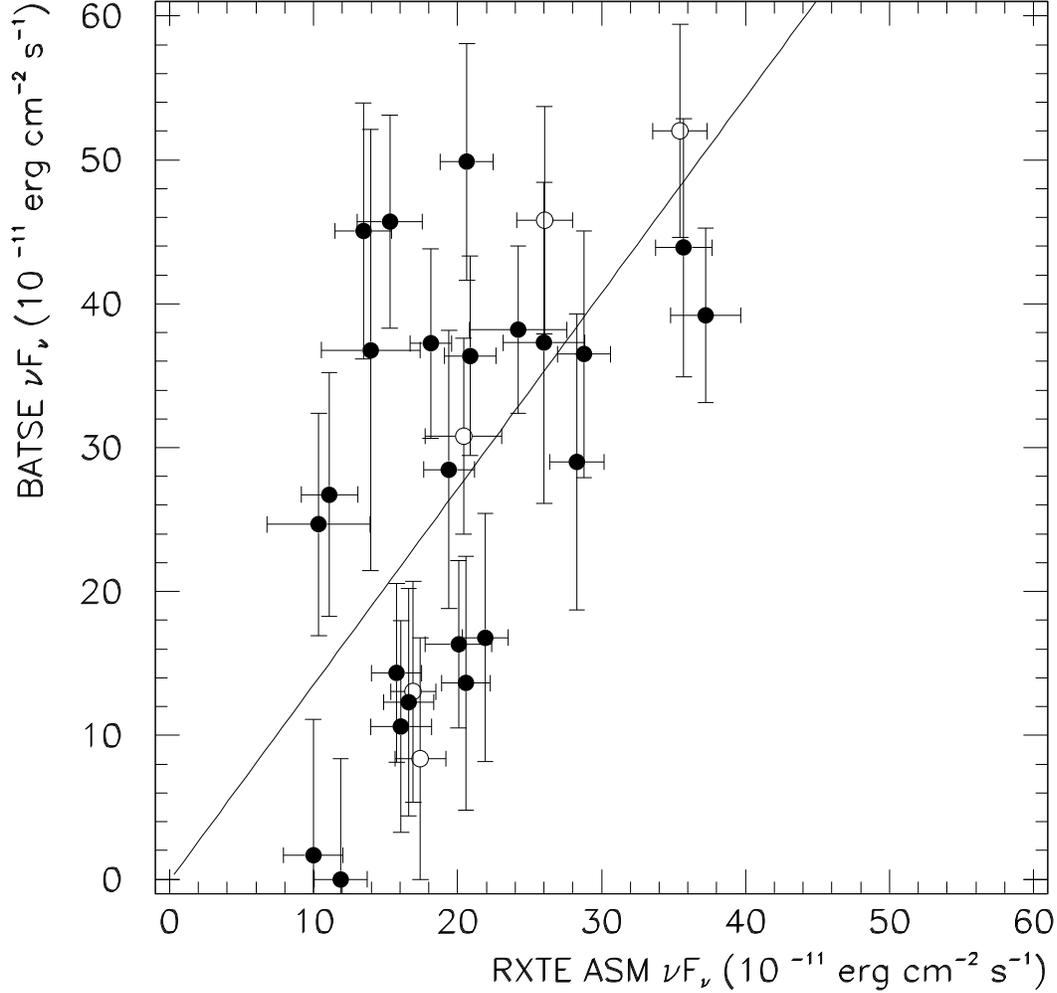}
\figcaption{\label{fig-batse-rxte}
The correlation between the BATSE hard X-ray and the RXTE ASM soft X-ray 
flux from Mkn~501. Here the time coverage is not systematically different,
only the BATSE duty cycle is much lower than that of the RXTE ASM. 
So all points are used in the linear fit. 
The points where HEGRA had bad time
coverage are are still marked as open circles for comparison with the
other figures. 
The fit of a linear function yields 
$y = (-0.1 \pm 5.3) + (1.36 \pm 0.24) x$, where
$y = $~BATSE flux at 36.4 keV in units of $10^{-11}$~erg~cm$^{-2}$~s$^{-1}$,
and $x = $~RXTE flux at 5.2 keV in the same units, and results in a reduced
$\chi^2$ of $2.3$. Only statistical errors were taken into account.
The linear correlation coefficient is $0.53$.}
\end{figure}

Figure \ref{fig-batse-rxte} shows the correlation between the weekly
Hard-X-ray and Soft-X-ray points. In this case, the time coverage is
the same for both instruments, only the duty cycles are different.
The correlation coefficient is 0.53 corresponding to a 0.5~\%
chance probability for a linear correlation. The constant term
of the linear fit is very well consistent with zero.
This figure also illustrates the difference in the dynamical ranges
of the variability in the soft and the hard X-ray band. At BATSE
energies, which are believed to be near the high-energy end of
the synchrotron spectrum, the variability amplitude is about 50 \% 
larger than at RXTE energies. This fits nicely into the scheme
that the strongest variability takes place at the high-energy
ends of both spectral components.

\subsection{Model fit results}

Each weekly SED is compared to a three-dimensional mesh of 
$48 \times 22 \times 9 = 9504$ simulations (48 different 
values of $p$, 22 values of $n_e$, and 9 values of $\gamma_2$), 
selecting the simulated broadband spectrum with the 
smallest $\chi^2$. The resulting best-fit parameters
are listed in table \ref{fit_parameters}, along with the
resulting $\chi^2$ divided by the number of data points. 
Since we do not have a continuous sequence in parameter 
space, the actual number of degrees of freedom is 
questionable so that we use the above quantity to
assess the quality of the fit. 

Our best-fit parameters for each individual weekly averaged
broadband spectrum are listed in Tab. \ref{fit_parameters}. 
With few exceptions (MJD~50549.0, 50576.9, 50618.8,
50696.1), all fits resulted in acceptable $\chi^2$ values.

We find a correlation between the electron injection spectral
index $p$ and the X-ray and high-energy $\gamma$-ray fluxes, while 
there is no obvious correlation with the total electron density 
and/or the high-energy cut-off $\gamma_2$ of the injected electron 
spectrum. There also appears to be a weak anti-correlation between 
the jet filling factor $f \propto (\Delta t_{rep})^{-1}$ and the 
X-ray and HEGRA fluxes. This could indicate that during states of 
relatively low activity, the fluxes are dominated by a quasi-steady 
component from a continuous jet, while in high-activity state the 
emission is dominated by more isolated, eruptive events. However, 
this latter correlation is much less pronounced than the correlation 
with the electron spectral index and will need to be tested on 
the basis of future, more sensitive observations.

\begin{figure}
\centering
\epsfysize=15cm
\rotatebox{270}{\epsffile{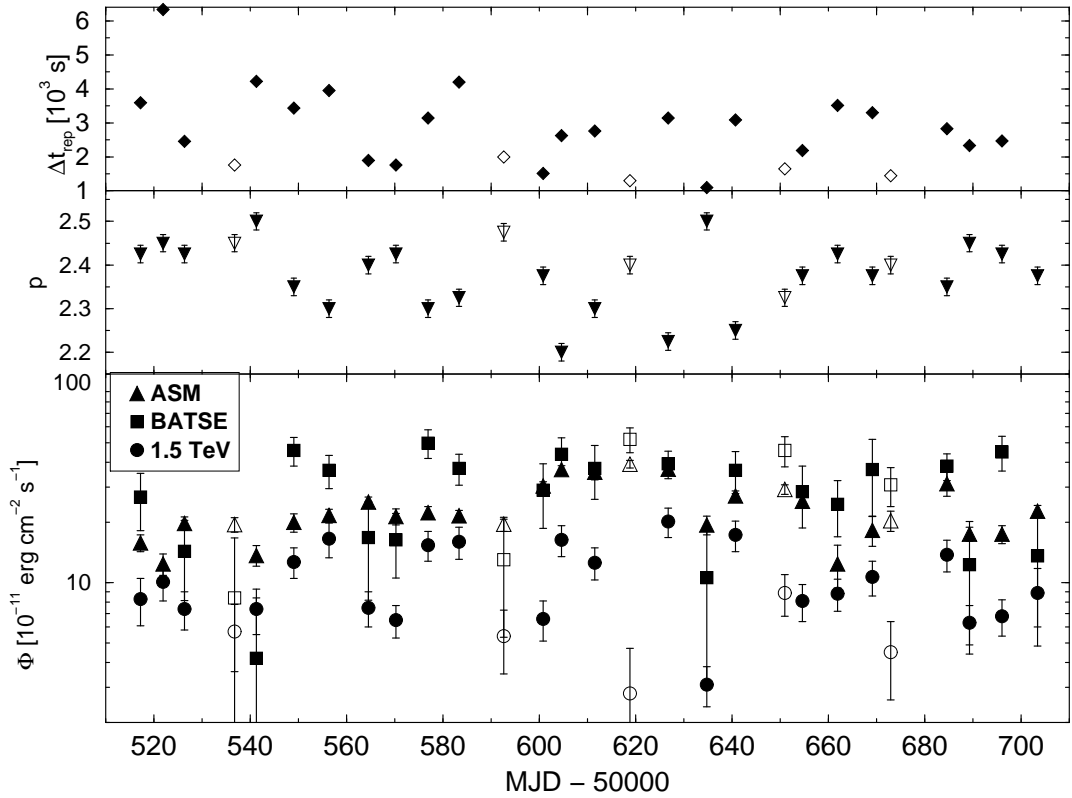}}
\figcaption{\label{par_corr}
Best-fit values of the blob ejection repetition time $\Delta t_{rep}$ 
and the electron injection spectral index $p$ for the pure SSC model 
compared to the RXTE ASM, BATSE and HEGRA 1.5 TeV light curves.} 
\end{figure}

\begin{figure}
\centering
\epsfxsize=14.5cm
\epsffile{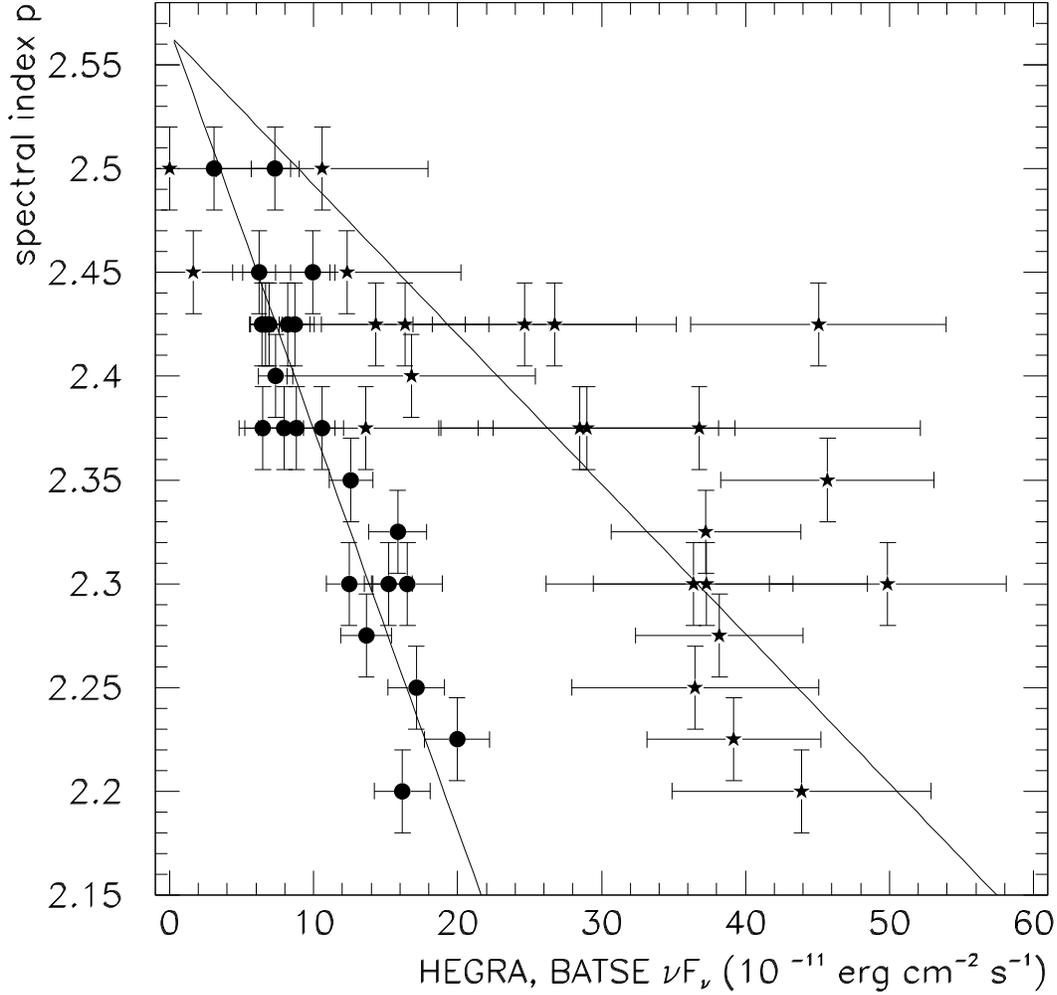}
\figcaption{\label{fig-p}
The correlation between the fit parameter spectral index, $p$, 
and the  HEGRA TeV (filled circles) and BATSE hard X-ray (stars) 
flux from Mkn~501. 
The fit of a linear function yields for HEGRA
$p = (2.57 \pm  0.018) + (-0.019 \pm 0.0018) x$, where
$x = $~HEGRA flux at 1.5 TeV in $10^{-11}$~erg~cm$^{-2}$~s$^{-1}$; 
reduced $\chi^2 = 1.18$; the linear correlation coefficient is $-0.89$.
For BATSE: $p = (2.56 \pm 0.034) + (-0.0072 \pm 0.0011) x$, where
$x = $~BATSE flux at 36.4 keV 
in $10^{-11}$~erg~cm$^{-2}$~s$^{-1}$; reduced $\chi^2 = 1.2$; 
the linear correlation coefficient is $-0.76$.
}
\end{figure}

In Fig. \ref{par_corr} the temporal variation of the best-fit 
values of $p$ and $\Delta t_{rep}$ are compared to the variations of 
the soft and hard X-ray fluxes and the 1.5 TeV flux. The correlation 
between the hard X-ray and TeV $\gamma$-ray fluxes with the 
injection spectral index is illustrated in figure 
\ref{fig-p}.

\begin{figure}
\centering
\epsfxsize=14.5cm
\epsffile{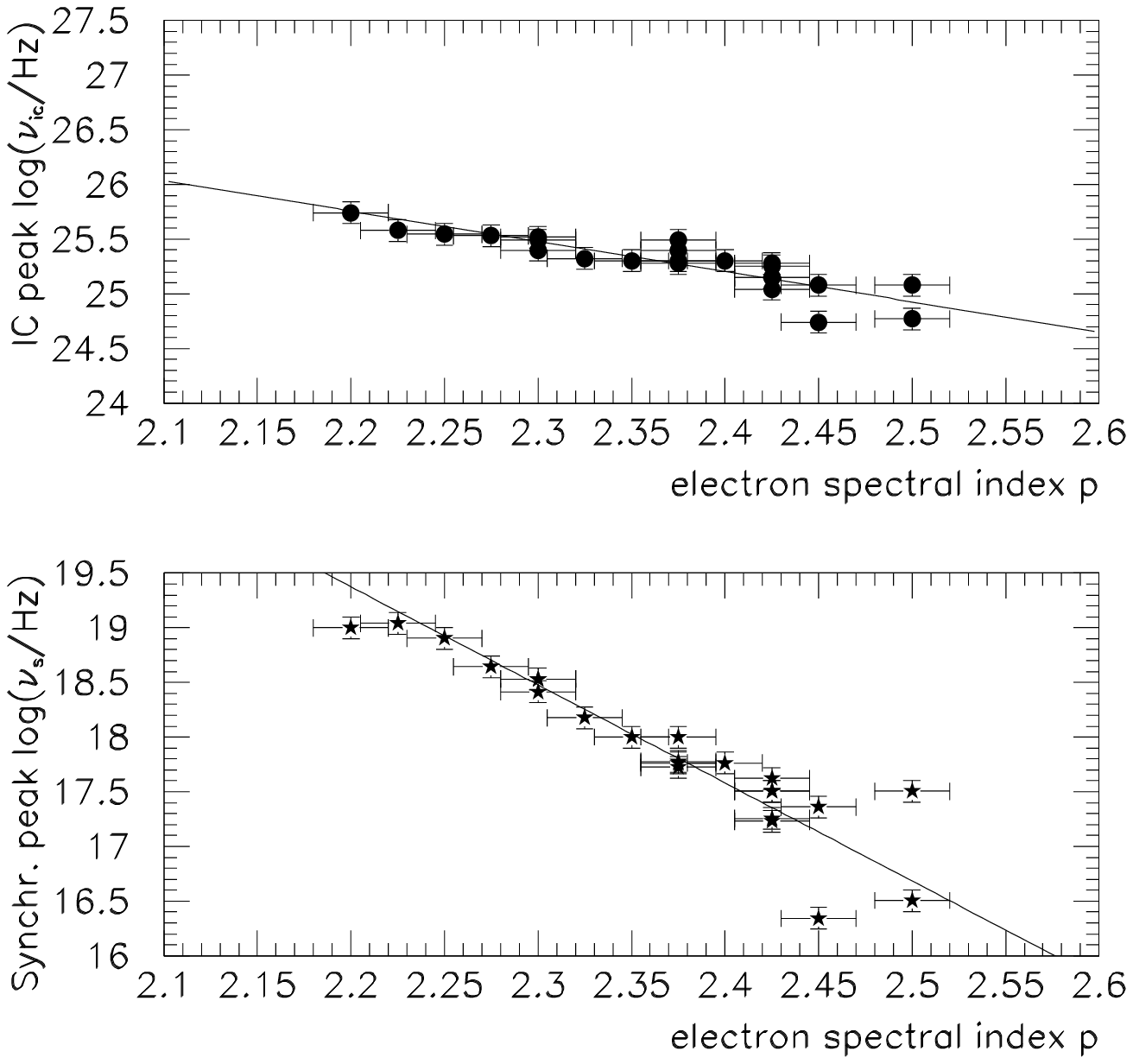}
\figcaption{\label{fig-peakpos}
The correlation between the fit parameter spectral index, $p$, 
and the  positions of the peak in the synchrotron (bottom plot) and 
inverse Compton (top plot) component of the weekly SEDs. 
The fit of a linear function yields 
$\log(\nu_{s}/\mathrm{Hz}) = (39.11 \pm  0.68) + (-8.97 \pm 0.29) p$ (linear 
correlation coefficient $-0.93$) and
$\log(\nu_{ic}/\mathrm{Hz}) = (31.84 \pm  0.38) + (-2.77 \pm 0.16) p$
(linear correlation coefficient $-0.88$).
}
\end{figure}

Furthermore, we show in figure \ref{fig-peakpos} the correlation between $p$ 
and the positions of the peaks in the
synchrotron and the inverse Compton component of the SED.
The peak positions are not fit parameters but were determined by finding the
local maxima in the weekly SEDs. We find that, in our model,  there is a 
strong correlation between the peak positions and $p$ such that these
parameters can be regarded as nearly identical. However, while the 
peak positions 
are directly observable, the electron spectral index is the more fundamental
quantity.

Our results indicate that medium-timescale high activity states in 
X-rays and high-energy $\gamma$-rays are consistent with a hardening 
of the electron spectrum injected at the base of the jet. As pointed 
out in the previous subsection, a pure SSC model in which the TeV
$\gamma$-ray flux is strongly influenced by Klein-Nishina effects,
predicts that the HEGRA and both the soft and hard X-ray fluxes should
roughly be correlated by  power-laws $F_{HEGRA} \propto F_X^{\delta}$
with $1.4 \lesssim \delta \lesssim 1.6$ in high flux-level states, in
which the contribution of a possible quasi-stationary radiation component
is small. The data available for this study are consistent with this but 
do not allow a clear distinction between different variability mechanisms, 
and future observations with increased sensitivity, in particular at multi-GeV
to TeV energies, are needed in order to test this prediction.

\section{Summary and Conclusions}

We have presented broadband spectra of the extreme HBL Mkn~501
during its high state in 1997, including radio, optical,
soft and hard X-ray, and TeV $\gamma$-ray observations. In this
study we concentrated on the medium-timescale variability, using
weekly averaged SEDs. We confirmed the strong correlation
between the TeV $\gamma$-ray flux and the hard X-ray flux. This
correlation was found to be non-linear and could be fitted with 
a second-order polynomial, in agreement with the expectation of 
an SSC dominated leptonic jet model, if the flux variations are 
related to fluctuations of the electron density in the jet and/or 
the spectral index of the electron spectrum at the time of injection 
into the jet.

The weekly averaged SEDs were fitted with a leptonic jet model,
strongly dominated by the SSC process. With a few exceptions,
this model yielded acceptable fits to the observed broadband
spectra. The observed spectral variability of Mkn~501 could be 
explained mainly by variations of the electron spectral index.
No clear correlation between the maximum electron energy
and the hard X-ray and TeV $\gamma$-ray fluxes on the 1-week
timescale was found, in contrast to the short-term variability 
of Mkn~501. Pian et al. (\cite{pian}) have shown that the
intraday variability of this object is most probably related
to an increase of $\gamma_2$, leading to pronounced flares in
hard X-rays, most probably on the synchrotron cooling timescale 
which is most likely $\sim$~a few hours and thus much shorter 
than the 1-week timescale considered in this paper. Our result
indicates that such synchrotron flares are isolated events and
are at most weakly correlated to the activity of the source
on the 1-week timescale.

Our result that the flaring behaviour on intermediate timescales
is consistent with a hardening of the electron spectrum is in
contrast to the flaring characteristics observed in quasars.
Recently, Mukherjee et al. (\cite{mukherjee99}) have investigated
all available broadband data on the very luminous flat-spectrum
radio quasar (FSRQ) PKS~0528+134, and found that its flaring 
behaviour is consistent with an increasing contribution of the 
external inverse-Compton component during flares, possibly related 
to an increase in the bulk Lorentz factor. The fits to the SEDs
of PKS~0528+134 required that the average energy of relativistic
electrons in the jet shifts towards lower values during flares, 
in contrast to the results found for Mkn~501. As pointed out by
B\"ottcher (\cite{boettcher99}), this implies that the synchrotron
peak is expected to shift towards lower frequencies during flares
of FSRQs, while Mkn~501 and Mkn~421 show clear evidence for a
shift of the synchrotron peak to higher frequencies. 

We point out that in the present study the magnetic 
field along the jet and the bulk Lorentz factor of
individual blobs were fixed, so that we cannot 
confidently rule out variations of the Doppler 
factor, accompanied by appropriate changes of the 
electron injection spectrum, as the flaring mechanism 
for Mkn 501. However, the very moderate variability
at optical frequencies, as observed in Mkn 501, leads
us to consider this flaring mechanism less likely in
this object since it would require a peculiar conspiracy 
between the Doppler factor and the electron spectrum to
keep the optical flux at an approximately constant level.

\section*{Acknowledgements}

The work of DP is supported by the Spanish CICYT grant SB97-B12601316. 
The work of MB was supported by NASA grant NAG 5-4055 (until Aug. 1999) 
and by Chandra Postdoctoral Fellowship grant number PF~9-10007, awarded 
by the Chandra X-ray Center, which is operated by the Smithsonian 
Astrophysical Observatory for NASA under contract NAS~8-39073.
The RXTE ASM data has been obtained through the High Energy Astrophysics
Science Archive Research Center Online Service provided by the
NASA/Goddard Space Flight Center. We thank C.D. Dermer for valuable 
comments on the manuscript and J.M. Holeczek for providing a fit routine.

\newpage

\begin{table}
\caption{\label{tab-instruments} The instruments which contributed 
data to this paper.}
\centering
\small
\begin{tabular}{lccc}
\hline
Instrument & \multicolumn{2}{c}{energy range} & comment \\
           & (Hz) &  (eV) &   \\ 
\hline
\rule[-1mm]{0cm}{0.5cm}Mets\"ahovi Radio Telescope &    
$22 \times 10^9$ & $9 \times 10^{-5}$ & $\lambda = $ 1.4 cm  \\
\hline
\rule{0cm}{0.4cm}Nordic Optical Telescope &      $4.4\times 10^{14}$ 
\--  $ 6.2 \times 10^{14}$   & 1.8 \-- 2.6  & 2.5 m mirror, filters: BVRI \\
\rule{0cm}{0.4cm}Tuorla Observatory &      -"-  & -"-   & 1.0 m mirror, filters: V \\
Osservatorio di Torino &     -"-  & -"-   & 1.0 m mirror, filters: BVR \\
\rule[-1mm]{0cm}{0.4cm}Osservatorio di Perugia &    
-"-  &  -"-  & 0.4 m mirror, filters: VRI\\
\hline
\rule[-1mm]{0cm}{0.5cm}RXTE ASM &    $4.8\times 10^{17}$  \--  $ 2.4 
\times 10^{18}$ & $2\times 10^{3}$  \--  $1\times 10^{4}$ &  (see e.g. 
Levine et al. 1996)\\
\hline
\rule[-1mm]{0cm}{0.5cm}BATSE &     $4.8\times 10^{18}$  \--  $ 4.8 \times 
10^{19}$ & $2\times 10^{4}$  \--  $2\times 10^{5}$ & occultation 
measurement\\
\hline
\rule{0cm}{0.4cm}HEGRA CT1 &     $3.6\times 10^{26}$ \--  $\approx 
7.3 \times 10^{27}$ & $1.5\times 10^{12}$  \--  $\approx 3\times 10^{13}$ &
  Aharonian et al. (1999b) \\
\rule[-1mm]{0cm}{0.4cm}HEGRA CT System &     $1.9\times 10^{26}$ \-- 
  $\approx 1.2 \times 10^{28}$ & $8\times 10^{11}$  \--   $\approx 
5\times 10^{13} $ &
  Aharonian et al. (1999a)\\
\hline
\end{tabular}
\end{table}

\begin{table}
\small
\caption{\label{tab-data} The weekly $\nu F_{\nu}$ datapoints from 
observations of Mkn 501 used for the model fits presented in this
paper. All entries are in units of $10^{-11}$ erg cm$^{-2}$ s$^{-1}$.  
Entries exactly equal to 0 are those where no data are available. 
This is mainly radio data. Upper limits are indicated by a ``$<$'' 
symbol and are calculated  with 90\% confidence level. The fluxes
at 0.8~TeV and 10~TeV are extrapolated from the flux measurement 
at 1.5~TeV.}
\begin{tabular}{llllrrrrr}
\hline
MJD & Radio & Optical & Soft X-ray & Hard X-ray & (0.8 TeV) & 1.5 TeV & (10 TeV) \\
\hline
50517.199   & $    0.02244 \pm 0.00104  $ & $  7.068  \pm 0.124 $ & $  11.1 \pm  3.1 $ & $ 26.73  \pm  8.5 $ & $   8.7  \pm   1.9 $ & $  8.2 \pm 1.9 $ & $    2.41  \pm   0.65 $ \\    
50521.906   & $    0.02387 \pm 0.00156  $ & $  7.148  \pm 0.069 $ & $  10.0 \pm  3.1 $ & $  1.67  \pm  9.5 $ & $  10.6  \pm   1.5 $ & $ 9.9 \pm 1.5 $ & $    2.94  \pm   0.61 $ \\   
50526.34    & $    0.02200 \pm 0.0011   $ & $  7.275  \pm 0.047 $ & $  15.8 \pm  3.3 $ & $ 14.33  \pm  6.2 $ & $   7.4  \pm   1.4 $ & $  6.9 \pm 1.3 $ & $    2.05  \pm   0.49 $ \\    
50536.715   & $    0.02420 \pm 0.0011   $ & $  7.394  \pm 0.120 $ & $  17.4 \pm  3.5 $ & $  8.38  \pm  8.4 $ & $   5.6  \pm   1.9 $ & $  5.3 \pm 1.9 $ & $    1.57  \pm   0.59 $ \\    
50541.262   & $    0.02237 \pm 0.00064  $ & $  7.607  \pm 0.077 $ & $  11.9 \pm  3.0 $ & $<  8.4          $ & $   7.8  \pm   1.7 $ & $  7.3 \pm 1.7 $ & $    2.17  \pm   0.58 $ \\   
50549.012   & $    0                    $ & $  7.769  \pm 0.063 $ & $  15.3 \pm  3.8 $ & $  45.7  \pm  7.4 $ & $  13.4  \pm   1.4 $ & $ 12.6 \pm 1.5 $ & $    3.72  \pm   0.69  $ \\    
50556.32    & $    0.02508 \pm 0.00198  $ & $  7.708  \pm 0.109 $ & $  20.9 \pm  3.9 $ & $  36.38 \pm  6.9 $ & $  17.6  \pm   2.4 $ & $ 16.5 \pm 2.4 $ & $    4.88  \pm   0.99  $ \\    
50564.566   & $    0.02382 \pm 0.00052  $ & $  7.948  \pm 0.093 $ & $  21.9 \pm  3.8 $ & $  16.79 \pm  8.6 $ & $   7.9  \pm   1.2 $ & $  7.4 \pm 1.2 $ & $    2.18  \pm   0.47 $ \\    
50570.254   & $    0.02420 \pm 0.00095  $ & $  7.941  \pm 0.075 $ & $  20.1 \pm  4.3 $ & $  16.34 \pm  5.8 $ & $   6.8  \pm   0.8 $ & $  6.4 \pm 0.8 $ & $    1.90  \pm   0.36 $ \\    
50576.934   & $    0.02409 \pm 0.00104  $ & $  8.071  \pm 0.188 $ & $  20.6 \pm  3.9 $ & $  49.87 \pm  8.2 $ & $  16.2  \pm   1.5 $ & $ 15.2 \pm 1.7 $ & $    4.49  \pm   0.80  $ \\    
50583.305   & $    0                    $ & $  7.921  \pm 0.317 $ & $  18.1 \pm  3.2 $ & $  37.24 \pm  6.6 $ & $  16.9  \pm   1.9 $ & $ 15.8 \pm 2.0 $ & $    4.68  \pm   0.89  $ \\    
50592.621   & $    0                    $ & $  7.829  \pm 0.153 $ & $  16.9 \pm  3.2 $ & $  13.03 \pm  7.7 $ & $   5.7  \pm   1.8 $ & $  5.3 \pm 1.7 $ & $    1.58  \pm   0.55 $ \\    
50600.797   & $    0                    $ & $  7.393  \pm 0.188 $ & $  28.3 \pm  4.7 $ & $  29.0  \pm 10.3 $ & $   6.9  \pm   1.3 $ & $  6.5 \pm 1.2 $ & $    1.92  \pm   0.46  $ \\    
50604.852   & $    0.02574 \pm 0.00132  $ & $  0                $ & $  35.7 \pm  5.6 $ & $  43.89 \pm  9.0 $ & $  17.2  \pm   1.8 $ & $ 16.2 \pm 1.9 $ & $    4.78  \pm   0.88  $ \\    
50611.523   & $    0                    $ & $  7.029  \pm 0.317 $ & $  26.0 \pm  5.4 $ & $  37.29 \pm  11 $ & $  13.3  \pm   1.5 $ & $ 12.5 \pm 1.6 $ & $    3.68  \pm   0.69  $ \\    
50618.848   & $    0                    $ & $  0                $ & $  35.5 \pm  5.4 $ & $  52.0  \pm  7.4  $ & $   3.0  \pm   2.0 $ & $  2.8 \pm 1.8 $ & $    0.83  \pm   0.56 $ \\    
50626.785   & $    0.02420 \pm  0.00154 $ & $  0                $ & $  37.2 \pm  6.2 $ & $  39.19 \pm  6.1  $ & $  20.7  \pm   2.1 $ & $ 20.0 \pm 2.3 $ & $    5.90  \pm   1.07  $ \\    
50634.707   & $    0.02189 \pm  0.00148 $ & $  7.327  \pm 0.070 $ & $  16.1 \pm  3.7 $ & $  10.61 \pm  7.4 $ & $   3.3  \pm   0.6 $ & $  3.1 \pm 0.5 $ & $    0.92  \pm   0.21 $ \\    
50640.742   & $    0                    $ & $  7.271  \pm 0.138 $ & $  28.8 \pm  4.7 $ & $  36.5  \pm  8.6 $ & $  18.2  \pm   1.8 $ & $ 17.1 \pm 2.0 $ & $    5.06  \pm   0.92  $ \\    
50650.969   & $    0                    $ & $  7.859  \pm 0.142 $ & $  26.0 \pm  4.5 $ & $  45.82 \pm  7.9 $ & $   9.4  \pm   1.8 $ & $  8.8 \pm 1.7 $ & $    2.60  \pm   0.63 $ \\    
50654.59    & $    0                    $ & $  0                $ & $  19.4 \pm  3.7 $ & $  28.48 \pm  9.7 $ & $   8.5  \pm   1.3 $ & $  8.0 \pm 1.3 $ & $    2.36  \pm   0.52 $ \\    
50661.887   & $    0.01672 \pm  0.00132 $ & $  7.226  \pm 0.317 $ & $  10.4 \pm  4.6 $ & $  24.66 \pm  7.7 $ & $   9.2  \pm   1.0 $ & $  8.7 \pm 1.1 $ & $    2.57  \pm   0.48  $ \\    
50669.164   & $    0.02706 \pm  0.00154 $ & $  7.567  \pm 0.217 $ & $  14.0 \pm  4.8 $ & $  36.78 \pm  15 $ & $  11.3  \pm   1.5 $ & $ 10.6 \pm 1.5 $ & $    3.12  \pm   0.63   $ \\    
50672.938   & $    0                    $ & $  7.244  \pm 0.160 $ & $  20.4 \pm  4.7 $ & $  30.82 \pm  6.8 $ & $   4.7  \pm   1.9 $ & $  4.5 \pm 1.8 $ & $    1.32  \pm   0.57 $ \\    
50684.598   & $    0                    $ & $  7.484  \pm 0.177 $ & $  24.2 \pm  5.8 $ & $  38.19 \pm  5.8 $ & $  14.6  \pm   1.7 $ & $ 13.7 \pm 1.8 $ & $    4.04  \pm   0.77  $ \\    
50689.281   & $    0                    $ & $  7.162  \pm 0.094 $ & $  16.6 \pm  3.4 $ & $  12.31 \pm  7.9 $ & $   6.6  \pm   1.1 $ & $  6.2 \pm 1.2 $ & $    1.84  \pm   0.42 $ \\    
50696.059   & $    0                    $ & $  0                $ & $  13.5 \pm  3.3 $ & $  45.06 \pm  8.9 $ & $   7.1  \pm   1.1 $ & $  6.7 \pm 1.1 $ & $    1.98  \pm   0.43 $ \\    
50703.445   & $    0                    $ & $  0                $ & $  20.6 \pm  3.7 $ & $  13.63 \pm  8.8 $ & $   9.4  \pm   2.8 $ & $  8.8 \pm 2.7 $ & $    2.61  \pm   0.87 $ \\
\hline
\end{tabular}
\end{table}    

\newpage

\begin{table}
\small
\caption{\label{fit_parameters} Best-fit parameters of a
pure SSC jet model to the weekly SEDs, and the quality of
the fit, indicated by $\chi^2$ / (no. of data points).
The $\chi^2$ is calculated without taking into account
the radio point. Periods marked with a ``*'' have less than 3
HEGRA points. $\nu_s$ and $\nu_{ic}$ are respectively the 
positions of the peak in the synchrotron and inverse compton component
of the SED. They were determined by finding the two local maxima
in each fitted SED. 
}
\begin{tabular}{lcccccccc}
\hline
MJD-50000 & $\gamma_2$/10$^7$ &  $p$  & $n_e$ & $\Delta t$ [10$^3$ s]  & $\log(\nu_{s}/\mathrm{Hz})$  & $\log(\nu_{ic}/\mathrm{Hz})$ & red. $\chi^2$ & no. data points \\
\hline
517.199   & 3.0   & 2.425 &  30 & 3.443 &  17.23    & 25.04  &   0.821  &    7 \\
521.906   & 3.0   & 2.450 &  65 & 6.189 &  16.34    & 24.74  &   0.395  &    7 \\
526.340   & 3.0   & 2.425 &  20 & 2.308 &  17.51    & 25.26  &   0.291  &    7 \\
536.715 * & 3.0   & 2.450 &  15 & 1.605 &  17.54    & 25.11  &   0.254  &    7 \\
541.262   & 3.0   & 2.500 &  50 & 4.071 &  16.51    & 24.77  &   0.539  &    7 \\
549.012   & 3.0   & 2.350 &  25 & 3.285 &  18.00    & 25.30  &   3.264  &    6 \\
556.320   & 3.0   & 2.300 &  25 & 3.798 &  18.53    & 25.40  &   0.806  &    7 \\
564.566   & 3.0   & 2.400 &  15 & 1.737 &  17.76    & 25.30  &   0.326  &    7 \\
570.254   & 3.0   & 2.425 &  15 & 1.614 &  17.62    & 25.28  &   0.495  &    7 \\
576.934   & 3.0   & 2.300 &  20 & 2.991 &  18.53    & 25.49  &   2.039  &    7 \\
583.305   & 3.0   & 2.325 &  30 & 4.048 &  18.18    & 25.32  &   1.760  &    6 \\
592.621 * & 3.0   & 2.475 &  20 & 1.849 &  17.34    & 25.08  &   0.241  &    6 \\
600.797   & 3.0   & 2.375 &  10 & 1.363 &  18.00    & 25.49  &   0.412  &    6 \\
604.582   & 2.5   & 2.200 &  10 & 2.476 &  19.00    & 25.74  &   0.309  &    6 \\
611.523   & 3.0   & 2.300 &  15 & 2.609 &  18.42    & 25.52  &   0.483  &    6 \\
618.848 * & 2.0   & 2.400 &  10 & 1.146 &  17.81    & 25.40  &   5.177  &    5 \\
626.785   & 3.0   & 2.225 &  15 & 2.989 &  19.04    & 25.58  &   0.140  &    6 \\
634.707   & 3.0   & 2.500 &  10 & 0.940 &  17.51    & 25.08  &   0.522  &    7 \\
640.742   & 3.0   & 2.250 &  15 & 2.941 &  18.90    & 25.54  &   0.354  &    6 \\
650.969 * & 3.0   & 2.325 &  10 & 1.500 &  18.23    & 25.53  &   1.326  &    6 \\
654.590   & 3.0   & 2.375 &  15 & 2.039 &  17.78    & 25.40  &   0.560  &    5 \\
661.887   & 3.0   & 2.425 &  30 & 3.361 &  17.26    & 25.15  &   0.959  &    7 \\
669.164   & 3.0   & 2.375 &  25 & 3.151 &  17.72    & 25.28  &   0.698  &    7 \\
672.938 * & 3.0   & 2.400 &  10 & 1.291 &  17.78    & 25.38  &   0.706  &    6 \\
684.598   & 3.0   & 2.275 &  15 & 2.676 &  18.64    & 25.53  &   0.849  &    6 \\
689.281   & 3.0   & 2.450 &  20 & 2.175 &  17.36    & 25.08  &   0.378  &    7 \\
696.059   & 3.0   & 2.425 &  20 & 2.319 &  17.51    & 25.15  &   2.996  &    5 \\
703.445   & 3.0   & 2.375 &  20 & 2.629 &  17.76    & 25.30  &   0.197  &    5 \\
\hline
\end{tabular}
\begin{tabular}{rcl}
\multicolumn{3}{l}{Fit parameters:}\\
         $\gamma_2$ & = & electron spectrum high-energy cutoff    \\
               $p$ & = & electron spectral index, $n(\gamma) \propto \gamma^{-p}$  \\
   $n_e$ [cm$^{-3}$] & = & electron density  \\
          $\Delta t$ [s] & = & blob ejection events repetition time scale (normalization)   \\
\multicolumn{3}{l}{Fixed parameters:}\\
               $z_i$ & = & 0.03 pc   \ \   (injection height of blob) \\
            $M_{BH}$ & = & $10^8 M_0$        \ \  (mass of central black hole) \\
               $L_D$ & = & $5 \times 10^{43}$ erg\,s$^{-1}$  \ \  (isotropic accretion disk luminosity)\\
               $R_B$ & = & $3 \times 10^{15}$ cm        \ \  (blob radius in the comoving frame) \\
           $\gamma_1$ & = & 200            \ \  (low-energy cutoff of electron sp.) \\
             $\Gamma$ & = & 25             \ \  (bulk Lorentz factor) \\
                 $D$ & = & 30             \ \  (Doppler factor) \\
    $\delta t_{min}$ & = & 3336 s   \ \  (contracted blob crossing time) \\
                 $B$ & = & 0.05 G         \ \  (magnetic field) \\
               $H_0$ & = & 75 km\,s$^{-1}$Mpc$^{-1}$  \ \  (Hubble constant) \\
               $q_0$ & = & 0.5            \ \  (deceleration parameter) \\
                 $z$ & = & 0.034          \ \  (cosmological redshift) \\
               $d_L$ & = & 133 Mpc        \ \  (luminosity distance) \\
\end{tabular}
\end{table}

\clearpage

\clearpage

\clearpage

\clearpage

\clearpage

\clearpage

\clearpage

\clearpage

\clearpage

\clearpage

\clearpage

\begin{thebibliography}{}

\bibitem[1999a]{hegraparti} Aharonian, F.A., Akhperjanian, A.G., Barrio, J.A. et al. 1999a, A\&A, 342, 69

\bibitem[1999b]{hegrapartii} Aharonian, F.A., Akhperjanian, A.G., Barrio, J.A. et al. 1999b, A\&A, 349, 29 
 
\bibitem[1999c]{hegraspec} Aharonian, F.A., Akhperjanian, A.G., Barrio, J.A. et al. 1999c, A\&A, 349, 11

\bibitem[1999]{ang99} Anguelov, V., Petrov, S., Gurdev, L., \& 
Kourtev, J., 1999, J. Phys. G.: Nucl. Part. Phys., 25, 1733

\bibitem[1998]{bednarek} Bednarek, W. 1998, A\&A, 336, 123

\bibitem[1999]{bednarek99} Bednarek, W., \& Protheroe, R. J. 1999,
MNRAS, in press (astro-ph/9902050)

\bibitem[1995]{blandford} Blandford, R. D., \& Levinson, A. 1995, ApJ, 
441, 79

\bibitem[1996]{bloom} Bloom, S. D., \& Marscher, A. P. 1996, ApJ, 461, 657

\bibitem[1997]{boettcher97} B\"ottcher, M., Mause, H., \& Schlickeiser, R.,
1997, A\&A, 324, 395

\bibitem[1998]{boettcher98} B\"ottcher, M., \& Dermer, C. D. 1998, ApJ, 
501, L51

\bibitem[1999]{boettcher99} B\"ottcher, M. 1999, ApJ, 515, L21

\bibitem[1997]{bradbury} 
        Bradbury, S.M., Deckers, T., Petry, D., et al. 1997, A\&{}A 320, L5

\bibitem[1997]{breslin} 
         Breslin, A.C., et al. 1997, IAU Circular No. 6592

\bibitem[1972]{colla} 
        Colla, G., et al. 1972, A\&{}AS 7, 1

\bibitem[1986]{crusius86} Crusius, A., \& Schlickeiser, R. 1986, A\&A, 164, L16

\bibitem[1992]{dermer92} Dermer, C. D., Schlickeiser, R., \& Mastichiadis, 
A. 1992, A\&A, 256, L27

\bibitem[1993]{dermer93} Dermer, C. D., \& Schlickeiser, R. 1993, ApJ, 
416, 458

\bibitem[1997]{dermer97} Dermer, C. D., Sturner, S. J., \& Schlickeiser, 
R. 1997, ApJS, 109, 103

\bibitem[1999]{cat} Djannati-Ata\"{\i}, A., Piron, F., Barrau, A., et al. 1998, submitted to A\&{}A (astro-ph/9906060)

\bibitem[1996]{fiorucci} Fiorucci, M. \& Tosti, G. 1996, A\&AS, 116, 403\\

\bibitem[1997]{fossati} Fossati, G., Celotti, A., Ghisellini, G., \& 
Maraschi, L. 1997, MNRAS, 299, 433

\bibitem[1996]{gaidos} Gaidos, J. A., Akerlof, C. W., Biller, S., et al.,
1996, Nature, 383, 319

\bibitem[1996]{ghisellini96} Ghisellini, G., \& Madau, P. 1996, MNRAS, 
280, 67

\bibitem[1998]{ghisellini98} Ghisellini, G., Celotti, A., Fossati, G., et 
al. 1998, MNRAS, 301, 451

\bibitem[1992]{batsemethod} Harmon, A., et al. 1992, AIP Conf. Proc. 280, 314

\bibitem[1998]{telarray} Hayashida, N., Hirasawa, H., Ishikawa, F., et al.,
1998, ApJ 504, L71

\bibitem[1999]{katajainen} Katajainen, S., Takalo, L.O., Sillanp\"a\"a, 
A., et al. 1999, A\&{}A, in press

\bibitem[1999]{kataoka} Kataoka, J., Mattox, J.R., Quinn, J., et al., 
1999, ApJ, 514, 138

\bibitem[1999]{kranich} Kranich, D., de Jager, O.C., Kestel, M. \& Lorenz, E., 1999, 
Proc. XXVI Int. Cosmic Ray Conf., Salt Lake City, OG 2.1.18   

\bibitem[1996]{levine} 
   Levine, A. M., Bradt, H., Cui, W., et al. 1996, ApJ, 469, L33 

\bibitem[1998]{malkan98} Malkan, M. A., \& Stecker, F. W. 1998, ApJ, 
496, 13

\bibitem[1993]{mannheim93} Mannheim, K. 1993, A\&A, 269, 67

\bibitem[1996]{mannheim96} Mannheim, K., Westerhoff, S., Meyer, H.,
\& Fink, H.-H. 1996, A\&A, 315, 77

\bibitem[1998]{mannheim98} Mannheim, K., 1998, Science, 279, 684

\bibitem[1992]{maraschi} Maraschi, L., Ghisellini, G., \& Celotti, A., 
1992, ApJ, 397, L5

\bibitem[1985]{marscher} Marscher, A. P., \& Gear, W. K. 1985, ApJ, 298, 
114

\bibitem[1994]{mastichiadis94} Mastichiadis, A., Protheroe, R. J. \& Szabo, A. P. 1994, MNRAS, 266, 910

\bibitem[1997]{mastichiadis97} Mastichiadis, A., \& Kirk, J. G 1997, 
A\&A, 320 19

\bibitem[1999]{mchardy} Mc Hardy, I. 1999, in ``BL Lac Phenomenon'', ASP Conference Series 159, 155

\bibitem[1999]{mukherjee99} Mukherjee, R., B\"ottcher, M., Hartman, R. C.,
et al. 1999, ApJ, 527, in press

\bibitem[1999]{nilsson} Nilsson K., et al. 1999,  submitted to AJ

\bibitem[1998]{pian} Pian, E., Vacanti, G., Tagliaferri, G., et al. 1998, ApJ, 492, L17

\bibitem[1981]{pravdo} 
   Pravdo, S.H. \& Serlemitsos, P. 1981, ApJ 246, 484

\bibitem[1997]{pravdo-b} 
   Pravdo, S.H., Angelini, L. \& Harding, A.K. 1997, ApJ 491, 808

\bibitem[1996]{quinn} 
        Quinn, J., Akerlof, C.W., Biller, S., et al. 1996, ApJ 456, L83

\bibitem[1999]{rachen99} Rachen, J.P. \& Mannheim, K. 1999, contribution
to the workshop ``On the astrophysics of relativistic sources'',
Marciana Marina, Elba, Italy

\bibitem[1999]{remillard} Remillard, R. 1999, RXTE help desk, private communication

\bibitem[1998]{whipple} Samuelson, F.W., Biller, S.D., Bond, I.H., et al. 1998, 
ApJ 501, L17

\bibitem[1978]{schwartz} 
        Schwartz, D.A., et al. 1978, ApJ 224, L103

\bibitem[1994]{sikora} Sikora, M., Begelman, M. C., \& Rees, M. J. 1994, 
ApJ, 421, 153

\bibitem[1998]{tav98} Tavecchio, F., Maraschi, L., \& Ghisellini, G.,
1998, ApJ, 509, 608

\bibitem[1998]{terasranta} Ter\"asranta, H., et al. 1998, A\&AS, 132, 305

\bibitem[1974]{toor} Toor, A. \& Seward, F.D. 1974, AJ 79, 995

\bibitem[1996]{tosti} Tosti, G., Pascolini, S., and Fiorucci, M. 1996, 
PASP, 108, 706

\bibitem[1975]{ulrich} 
        Ulrich, M.H., et al. 1975, ApJ 198, 261

\bibitem[1997]{villata97} Villata, M., Raiteri, C.M., Ghisellini, G., et 
al. 1997, A\&AS, 121, 119

\bibitem[1998]{villata98} Villata, M., Raiteri, C.M., et al. 1998, A\&AS, 
130, 305

\bibitem[1998]{villata99} Villata, M. \& Raiteri, C.M. 1999, A\&A, 
347, 30

\end{thebibliography}
\end{document}